\documentclass{aa}
\usepackage[authoryear]{natbib}
\usepackage{graphicx}
\usepackage{txfonts}
\usepackage[utf8]{inputenc}

\newcommand{\feh}{\mbox{[Fe/H]}}

\newcommand{\afe}{\mbox{[$\alpha$/Fe]}}
\newcommand{\cfe}{\mbox{[C/Fe]}}
\newcommand{\nfe}{\mbox{[N/Fe]}}
\newcommand{\nife}{\mbox{[Ni/Fe]}}
\newcommand{\tife}{\mbox{[Ti/Fe]}}
\newcommand{\cafe}{\mbox{[Ca/Fe]}}
\newcommand{\crfe}{\mbox{[Cr/Fe]}}
\newcommand{\cofe}{\mbox{[Co/Fe]}}
\newcommand{\mgfe}{\mbox{[Mg/Fe]}}
\newcommand{\mnfe}{\mbox{[Mn/Fe]}}
\usepackage{marginnote}
\usepackage{color}
\begin{document}

\title{Gemini spectroscopy of the outer disk star cluster BH176}
\titlerunning{Distant Galactic cluster BH176}

\author{M. E. Sharina \inst{1}\thanks{
Based on observations obtained at the Gemini Observatory, which is operated 
by the Association of Universities for Research in Astronomy, Inc., under a 
cooperative agreement with the NSF on behalf of the Gemini partnership: 
the National Science Foundation (United States), the Science and Technology Facilities
Council (United Kingdom), the National Research Council (Canada),
CONICYT (Chile), the Australian Research Council (Australia),
Minist\'{e}rio da Ci\^{e}ncia, Tecnologia e Inova\c{c}\~{a}o (Brazil),
and the Ministerio de Ciencia, Tecnolog\'{i}a e Innovaci\'{o}n Productiva
(Argentina).}
\and
C. J. Donzelli \inst{2}
\and
 E. Davoust \inst{3}
\and
V. V. Shimansky \inst{4}
\and
C. Charbonnel \inst{5}
}
\authorrunning{Sharina et al.~(2013)}
  \institute{Special Astrophysical Observatory, Russian Academy  of Sciences, N. Arkhyz, KChR, 369167, Russia\\
              \email{sme@sao.ru}
\and Instituto de Astronom\'ia Te\'orica y Experimental IATE, CONICET - Observatorio Astron\'omico, 
Universidad Nacional de C\'ordoba, Laprida 854, X5000BGR, C\'ordoba, Argentina \\
              \email{charly@oac.uncor.edu}
 \and IRAP, Universit\'e de Toulouse, CNRS, 14 avenue E. Belin, F-31400, France\\
              \email{edavoust@irap.omp.eu}
\and Kazan Federal University, Kremlevskaya 18, Kazan, 420008, Russia \\
\email{Slava.Shimansky@kpfu.ru} 
\and Geneva Observatory, University of Geneva, 51, ch. des Maillettes, 1290 Versoix, Switzerland\\
\email{corinne.charbonnel@unige.ch}}  
\date{Received August 15, 2013; accepted }

\abstract
{BH176 is an old metal-rich star cluster. It is spatially
and kinematically consistent with belonging to the Monoceros Ring.
It is larger in size and more distant from the Galactic plane than typical 
open clusters, and it does not belong to the Galactic bulge.}
{Our aim is to determine the origin of this unique object 
by accurately determining its distance, metallicity, and age. 
The best way to reach this goal is to combine spectroscopic and photometric methods.}
{We present medium-resolution observations of red clump and red giant branch stars 
in BH176 obtained with the Gemini South Multi-Object Spectrograph. 
 We derive radial velocities, metallicities, effective temperatures, and surface 
gravities of the observed stars and use these parameters to distinguish member stars
 from field objects.}
{We determine the following parameters for BH176: $V_h\!=\!0\pm 15$ km/s, $\feh\!=\!-0.1\pm 0.1$, 
age $7\pm 0.5$ Gyr, $E(V-I)\!=\!0.79\pm 0.03$, distance $ 15.2\pm 0.2$ kpc,
$\alpha$-element abundance $\afe \sim 0.25$ dex (the mean of [Mg/Fe], and [Ca/Fe]).}
{BH176 is a member of old Galactic open clusters that presumably belong 
to the thick disk. It may have originated as a massive star cluster 
after the encounter of the forming thin disk 
with a high-velocity gas cloud or as a satellite dwarf~galaxy.}


\maketitle


\section{Introduction}

Star clusters (SCs) are important tracers of the evolutionary history of our Galaxy.
The knowledge of their chemical, structural, and dynamical properties is indispensable for
understanding how the Galaxy formed and how much dwarf satellites contributed to its formation.
The age, distance, and metal content of SCs can be determined more easily than for single stars.
SCs are prime targets for improving our understanding of stellar evolution using 
deep colour-magnitude diagrams (CMD) and high-resolution spectroscopy. The old globular clusters (GCs),
which populate galactic haloes, 
were enriched in $\alpha$-elements during their formation due to explosions of SNeII
(Kruijssen \& Diederik and references therein). 
The high-metallicity GCs ($\feh > -0.4$ dex) belong to the bulge or the thick disk, but
their origin is still an open subject.  

Open clusters (OCs) reside in the thin disk. 
They are generally less massive and more metal-rich than GCs and lose stars
from tidal interactions with the interstellar medium. 
Because of the enormous extinction and density of matter near the Galactic plane 
and because of their loose structure, OCs have only been studied within a radius 
of $R_{gc}\sim$8 kpc from the Sun (Kharchenko et al. 2013). 
Young embedded SCs presently undergoing star formation have been discovered 
in dense molecular clouds.
Understanding the mechanisms triggering the formation of massive SCs is one of the primary
goals of contemporary astrophysics.

The nature of BH176 ($\alpha=15^h39^m07.3$, $\delta=-50\degr 03'02\arcsec$, J2000.0), 
which was discovered and classified as an OC by  van den Bergh \& Hagen (1975), 
is still a matter of debate.
The stellar populations of the SC were studied by photometric  
(Ortolani et al. 1995, Phelps \& Schick 2003) and by photometric and spectroscopic methods 
(Frinchaboy et al. 2006). The latter provided the first reliable radial
velocity ($ V_h =11.2 \pm 5.3$~km s$^{-1}$), distance from the Sun (15.8 kpc), 
photometric metallicity ($\feh \sim 0.0$ dex), and age (7 Gyr) for BH176.
Their conclusion was that the SC is a possible  member of the 
Monoceros stream according to its location and kinematics; however,
the high metallicity suggests that this connection is unlikely.
The metallicity distribution of the Monoceros stream was measured
by Meisner et al. (2012) from the spectroscopy of 594 stars 
obtained with the low-resolution MMT Hectospec. They found the stream to be chemically 
distinct from both the thick disk and halo, with $\feh = -1$~dex and a small 
intrinsic dispersion of $0.10 \div0.22$ dex. BH176 does not fit the age-metallicity relation of the Sgr dwarf spheroidal galaxy 
(Layden \& Sarajedini, 2000). Harris (1996) classified this SC 
as a GC with a distance of 15.6 kpc from the Sun, extinction $E(B-V)=0.77$ mag., 
visual level of the horizontal branch $V_{HB}=19.0$ mag., and total visual absolute 
magnitude $M_V=-4.35$ mag. 

Davoust, Sharina \& Donzelli (2011, hereafter D11) used another approach to redetermine the
age and metallicity of this unique object and to accurately derive its distance and Galactic extinction
using the Two Micron All Sky Survey (2MASS) Point Source catalogue and archival VLT images. 
Red giant branch (RGB) stars belonging to BH176 were detected independently
in the visual and infrared wavelength ranges in the region of the CMDs 
defined by the following equations:
 $ K_{s0}< -9.09(J-K_s)_0 + 20.0$, $ I< -6.9(V-I)+29,$ and $V-I>1.8$ (where V and I magnitudes 
are not corrected for Galactic extinction).
The spatial distribution of stars selected in this way allowed the authors 
to derive a maximum extent of 3 $\arcmin$ for the cluster.

Data from the 2MASS Point Source Catalog with a good photometric quality allowed them
to reach a photometric limit of $\sim$5~mag below the tip of the red giant branch (TRGB) in the $K_s$ band. 
The strong Galactic extinction was not an obstacle in the infrared; 
 however, owing to the poor photometric depth and the large number of foreground stars, 
it was difficult to derive the distance accurately. 
There are only three red giant stars within the boundaries of BH176 
$\sim 3$ mag. from the TRGB  ($K_s$=[12$\div$9]).
The derived extinction in the $K_s$ band was $A_{K_s}$=0.212$\pm 0.03$ mag.
Artificial star tests were made in the course of the stellar photometry on the VLT images 
to evaluate how the crowding influences the photometric depth and accuracy.
The deep visual CMD reached $\sim$1~mag. below the main sequence turn-off (MSTO) of the SC.
However, owing to strong contamination problems the ridge line outlining 
the upper MS, SGB, and RGB of the SC was defined only statistically.
Since the MSTO was close to the photometric limit and strongly contaminated by
foreground stars and because of the scarcity of stars on the TRGB,
only RC stars provided a good distance indicator representing a strong peak in the
colour-magnitude density distribution. 

Using theoretical predictions 
for the dependence of the I-band magnitude and $ (V-I)_0$ colour of the RC on the age and metallicity (Girardi \& Salaris, 2001), 
 the following parameters were obtained for BH176:
$ \feh=-0.17^{+0.08}_{-0.11}$~dex\footnote{$\feh=log(Fe/H)_{*} -log(Fe/H)_{\sun}$, 
where $log(Fe)= log(n(Fe)/n(H))+12$ and $n(Fe)$ and $n(H)$ are the numerical densities 
(cm$^{-3}$) of iron and hydrogen atoms, respectively.}, $ M_V (RC)= 0.88 \pm 0.07$, $ M_I (RC)=-0.17 \pm 0.07$, and 
$(m-M)_0 = 15.9 \pm0.07$. Combining these data with the ones derived using 
different methods (isochrone fitting, RGB bump luminosity) 
led to the following results: Dist=$ 15.1 \pm 0.5$ kpc, E(V-I) = 0.796, $ \feh=-0.1 \pm 0.1$ dex, 
age$=7$ Gyr. 
The authors suggested that BH176 is an open, or {\it transitional} cluster.
We simply use the term "star cluster" throughout this paper.
Indeed, BH176 has properties that are distinct from both OCs and GCs, but
the term "{\it transitional} cluster" has not been clearly defined.

In this study we present and analyse data from the Gemini South Multi-Object Spectrograph 
for RGB and RC stars in BH176, selected using the above photometric study (D11).
The pre-imaging run and spectroscopic observations are described in Sec.~2. 
The reduction method and spectroscopic results are summarised in Sec.~3.
A short Sec.~4 is devoted to the proper-motion issues.
In Sec.~5 isochrones are fit to the VLT photometric data
to improve the extinction, distance, and age estimates. The results are
discussed in Sec.~6 and summarised in Sec.~7.

\section{Observations and data reduction}

\begin{table}[!h]
\caption{Journal of pre-imaging and spectroscopic observations.} 
\label{tab:obslog}
\scriptsize
\begin{center}
\begin{tabular}{lccccr}
\\ \hline\hline\noalign{\smallskip}
Object &  Date & Filter/grism & $T_{exp}$ & Seeing  & Air mass\\
       &       &  &  (s)      & (arcsec)&          \\
\hline
pre image  & 2011-02-01  & G0325      & 1x60   & 0.95 & 1.39 \\
mask 1 & 2011-03-29  & B600-G5323 & 3x1200 & 0.94 & 1.09 \\
mask 2 & 2011-03-30  & B600-G5323 & 2x1200 & 0.98 & 1.14 \\
mask 2 & 2011-04-02  & B600-G5323 & 1x1200 & 0.98 & 1.16 \\
\noalign{\smallskip}
\hline
\end{tabular}
\end{center}
\end{table}

The spectroscopic observations were carried out with the Gemini South telescope using the
Gemini Multi Object Spectrograph (GMOS) as part of Gemini programme GS-2011A-Q-22 (PI: Donzelli).
The targets selected within the boundaries of BH176 generally follow 
the ridge line outlined by the VLT stellar photometry of D11. 
These are bright stars falling on the possible positions of the slits 
with magnitudes and colours roughly corresponding to the RC and RGB of the SC.
To use all the slits, some field stars with $V-I <1.8$ were also observed.

\subsection{Images}
To set the slits accurately (see next section) a 60-sec exposure
in the SDSS $g'$ filter was obtained (see Table~\ref{tab:obslog} for details). 
The field covers a region of 5 x 5 arcmin$^2$,
with a pixel scale of 0.145 arcsec, which is perfectly suitable for our targets. 
The image was provided by the Gemini staff, and it was automatically reduced by the Gemini pipeline. 
Two GMOS multi-slit
masks were designed with the GMOS Mask Making software using this preliminary image. 
The image was also used to transform the VLT coordinates of the target stars into the original 
astrometric system of the Gemimi South telescope.

\subsection{Spectra}
Using the B600-G5323 grating blazed at 5000 $\AA$, a slit width of 1 arcsec and a minimum slit
length of 4 arcsec, we were able to allocate 41 and 39 slits in masks 1 and 2, respectively.
In each case, a total of 3x1200 sec exposures were obtained at three different central wavelengths 
(4970, 5020, and 5070 $\AA$) in order to correct for the gaps between CCDs. 
A binning of 2x2 was used, yielding a scale of 0.146 arcsec pixel$^{-1}$, 
a theoretical dispersion of $\sim$ 0.9 $\AA$ per pixel, and a resolution of $FWHM\sim$ 5 $\AA$. 
The spectroscopic data were acquired in queue mode under very good seeing conditions
 (see Table~\ref{tab:obslog} for details).

The spectra cover the range  3600-6300 $\AA$, but the exact range 
depends on the slit position on the mask. 
Bias frames, dome, and sky flat fields and Copper-Argon (CuAr) arcs were taken for calibration.
Wavelength calibration left residuals of 0.2 $\AA$. 
Flux calibration was performed using the spectra of the standard star $LTT 9239$, 
which was obtained with the same instrument configuration as for the science
objects. All science and calibration files were retrieved from the Gemini Science Archive 
hosted by the Canadian Astronomy Data Centre. 

The data were reduced with the Gemini/GMOS package within IRAF following the standard procedure.
We normalised the spectroscopic flat fields with {\sc gsflat}, and then mosaicked the three GMOS CCDs
into one image with {\sc gmosaic}. We subtracted bias frames and flat-fielded the mosaic object frames using the task {\sc gsreduce}. 
In the case of the arc frames, we only did the bias subtraction. The procedure {\sc gswavelength} was used to determine 
the wavelength calibration from the CuAr spectra. Then the object frames were 
wavelength-calibrated using the task {\sc gstransform}. Once the object frames were calibrated we did 
the sky subtraction using {\sc gsskysub}. The one-dimensional individual spectra were extracted using {\sc gsextract},
and the final flux calibration was done with the {\sc gscalibrate} routine. Prior to the flux calibration,
the task {\sc gsstandard} was used to derive the sensitivity function.

\begin{figure}[!hbt]
\resizebox{\hsize}{!}{\includegraphics[width=6.7cm, angle=90]{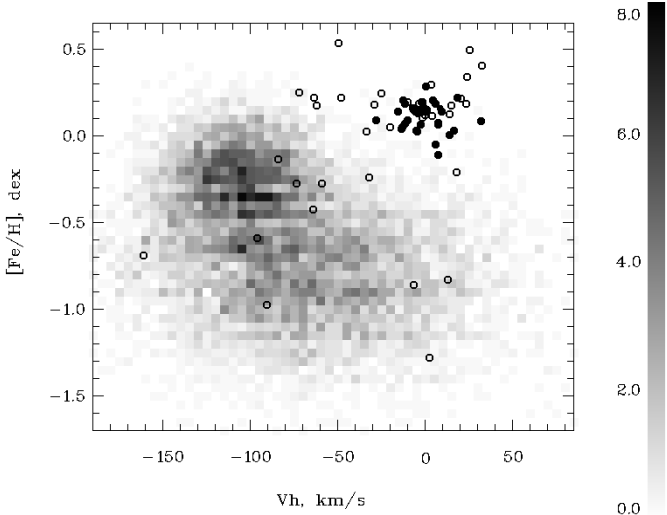}}
 \caption{Distribution in metallicity and heliocentric velocity (Table~\ref{tab:summary}) of all the stars observed spectroscopically (dots),
compared with the density distribution of the same parameters obtained using the Besan\c{c}on model.
Black dots show the stars classified as belonging to BH176.}
 \label{fig:Vh_Fe_H}
\end{figure}

\section{Analysis of the spectroscopic data}


To identify stars belonging to BH176 among our sample,
 we inspected theoretical and empirical stellar libraries for possible analogues.
The main disadvantage of empirical stellar libraries (and stellar population models based on them)
is that the abundance patterns are biased towards the solar neighbourhood and low-mass stars.
Theoretical spectra are not limited in atmospheric parameters, wavelength range,
spectral resolution, or abundance patterns. However, their computation is based on incomplete 
lists of atomic and molecular lines and inaccurate atomic constants 
(see e.g. Coelho 2014, Husser et al. 2013 for a thorough discussion).

Preliminary effective temperatures for the stars were estimated using the 
$(V-K_{s})_0$ colour indices and the temperature calibrations by Alonso et al. (1999). 
These data are listed in column 7 of Table~\ref{tab:summary}.
The successive steps of our spectroscopic analysis were the following.
First, we compared spectra of the observed stars in and around BH176 with the ELODIE 
stellar library (Prugniel \& Soubiran 2001, Prugniel et al. 2007). 
It provides a wide coverage of atmospheric parameters: $Teff=3100K \div 50000$~K, 
$log~g= -0.25\div 4.9$, and $\feh=-3 \div 1$~dex.
Most stars in this library have solar, or near solar, metallicity,
close to the one of BH176 (D11). 
Second, using the effective temperatures, surface gravities, metallicities, and radial velocities 
derived in the previous step, we developed criteria 
to separate stars belonging to BH176 from the field ones. 
Then we co-added spectra of the selected objects to obtain a
representative spectrum of high S/N, which was compared with 1) individual stars with similar
physical parameters in the MILES stellar library (S\'anchez-Bl\'azquez et al. 2006, Cenarro et al. 2007) and  
 2) synthetic spectra calculated from stellar atmosphere models
with various chemical compositions corresponding to the predefined atmospheric parameters.

\subsection{Full spectrum fitting in {\sc ULySS}}
\label{sec_ulyss}

To derive the properties of the stars observed in BH176 we used
the {\sc ULySS} program\footnote{http://ulyss.univ-lyon1.fr}
(see e.g. Koleva et al. 2009 for a detailed explanation of the software) 
with the interpolating arrow built on the ELODIE 3.2 stellar 
library\footnote{ http://www.obs.u-bordeaux1.fr/m2a/soubiran/elodie$_{-}$library.html; 
http://ulyss.univ-lyon1.fr/models.html }. The {\sc ULySS} program performs 
a non-linear least-squares minimisation of the difference between the model and observed
spectra, together with a normalisation of the pseudo-continuum and taking the 
line-spread function of the spectrograph
into account. Multiplicative and additive polynomials 
were applied to the observed spectrum to bring it in agreement with the model spectrum. 
Prugniel et al. (2011) and Wu et al. (2011) determined the atmospheric parameters
 T$_{eff}$, $ log(g)$ and $[Fe/H]$ for stars in the MILES and Coudé-feed Indo-US spectral libraries
 with this method.

 Before the fit we derived line-spread functions for each of the observed slits using twilight spectra
obtained at the same positions in the field of view.
The reference spectra were degraded to the resolution of the observed spectra.

The derived stellar parameters are listed in Table~\ref{tab:summary}. 
The columns contain the following information:
(1) slit number (a stands for mask 1, b stands for mask 2); 
(2) equatorial coordinate (J2000.0);
(3, 4) visual magnitude in the $I$-band and $(V-I)$ colour from D11, uncorrected 
for Galactic extinction;
(5, 6) $K_s$ magnitude and $(J-K_s)$ colour from the 2MASS Point Source Catalog;
(7) T$_{eff}$, in Kelvin, calculated from $(V-K)_0$ according to the empirical calibration of Alonso et al.(1999);
(8) distance from the centre of the SC, in arcsec; 
(9) heliocentric radial velocity with uncertainty, in km~s$^{-1}$; 
(10) metallicity $ [Fe/H]$, in dex; 
(11) effective temperature, in Kelvin, estimated in {\sc ULySS}; 
(12) surface gravity, in cm/s$^2$; 
(13, 14) proper motion in right ascension $\mu_{\alpha}*cos(\delta)$ and declination $ \mu_{\delta}$ from the 
PPMXL catalogue (Roeser et al. 2010), in mas yr$^{-1}$; 
(15) S/N per resolution element at 5000Å in the reduced one-dimensional spectrum;
(16) note on the stellar type.

The following information is collected in the last column of Table~\ref{tab:summary}.
The stars belonging to BH176 are marked by an asterisk (RC), or a double  
asterisk in the case of RGB stars. Three stars were classified as semi-regular variables 
(SRb) (see Section 3.3). TiO molecular bands were detected in some RGB stars.
The MgH molecular bands were much stronger in three field stars in comparison to other sample objects. 
Some objects were difficult to classify due to insufficient S/N or to narrow spectral ranges. They are marked by '?'.

\subsection{Identying stars belonging to BH176}

The distribution of the derived values of $V_h$ and $\feh$ for all the stars observed 
spectroscopically is shown in Fig.~\ref{fig:Vh_Fe_H} and compared 
with the density distribution of the corresponding data calculated using the 
Besan\c{c}on model of our Galaxy\footnote{http://model.obs$-$besancon.fr/modele$_{-}$options.html} 
(Robin et al. 2003, 2004). It also shows the stars classified as belonging to BH176
(marked by asterisks in Table~\ref{tab:summary}).
The following parameters were chosen to generate a 
catalogue of pseudo-stars in the Johnson-Cousins photometric system with the Besan\c{c}on model:
a distance interval 14.8 -- 16 (kpc), an area 1 deg$^2$ centred at 
 $l=328\degr$ and $b=4\degr$ in Galactic coordinates; magnitudes in the range 
$23\ge V\ge15$ and $21\ge I\ge13$; $-200\ge V_h\ge100$. 
The diffuse extinction parameter was set equal to 0.7 mag~kpc$^{-1}$.
The default values were adopted for the other parameters (spectral and luminosity types, 
photometric uncertainties, proper motions).  We then selected the stars with $-1.7\ge\feh\ge0.7$.
The resulting sample contained 9635 stars. Figure~\ref{fig:Vh_Fe_H} shows the
distribution of this artificial sample in heliocentric radial velocity and 
metallicity in the form of numbers of stars per unit bin with a size of
 $V_h=5$ km s$^{-1}$ and  $\feh=0.05$ dex.
If we select
as small an area as possible to fit the position and the size of BH176,
the $V_h$ and $\feh$ distribution generated by the Besan\c{c}on model
does not change significantly. 
We used an area that is large enough to obtain a statistically representative sample.

Figure\ref{fig:Vh_Fe_H} shows that this region of the Galaxy contains a negligibly small
number of stars with metallicities and radial
velocities similar to those of member stars of BH176.
This and the atmospheric parameters summarised in Table~\ref{tab:summary}
allow us to {\it identify the stars belonging to the SC.}
These are the {\it criteria}: 
$1.8\ge V-I\ge5$ mag., $13.87\ge I\ge 17.1$, $V_h\ge-40$ km~s$^{-1}$, 
$\feh\ge-0.24$ dex, $3100\ge T_{eff}\ge4900$ K, and $-0.5\ge$log~g$\ge3.5$. 

\subsection{A representative spectrum of RC stars in BH176}

\begin{table*}[!hbt]
\caption{Elemental abundances and
their uncertainties for the summed spectra of 31 RC stars in BH176 derived in this work 
using the stellar atmosphere models and
literature abundances of a similar star HD184406.}
\label{tab:abund}
\begin{center}
\begin{tabular}{|lccccccccc|}
\hline\hline 
Object  &  \cfe   & \nfe     & \mgfe & \cafe & \tife &  \crfe  & \mnfe  &  \cofe  & \nife \\
        & (dex)    & (dex)    & (dex)    & (dex)    & (dex)    & (dex)    & (dex)    & (dex)    & (dex)    \\
\hline
BH176            &  0.15 & 0.15 & 0.25&  0.25& 0.35& 0.25& -0.05&   0.25 & -0.05\\
      &$\pm$0.15& $\pm$0.20&  $\pm$0.10&  $\pm$0.15&  $\pm$0.20& $\pm$0.20 & $\pm$0.20 &  $\pm$0.20 &  $\pm$0.20 \\
\noalign{\smallskip}
HD184406 & 0.22$^1$ & 0.03$^1$& 0.35$^1$& 0.13$^2$(-0.49$^1$)& -0.16$^2$(-0.27$^1$)& -0.14$^1$&  -- & 0.11$^1$ & -0.21$^2$(0.26$^1$)\\
\hline
\end{tabular}
\tablefoot{  
The superscripts refer to (1) Luck \& Challener (1995), (2) McWilliam (1990).}
\end{center}
\end{table*}

\begin{figure}[!h]
\resizebox{\hsize}{!}{\includegraphics[angle=-90]{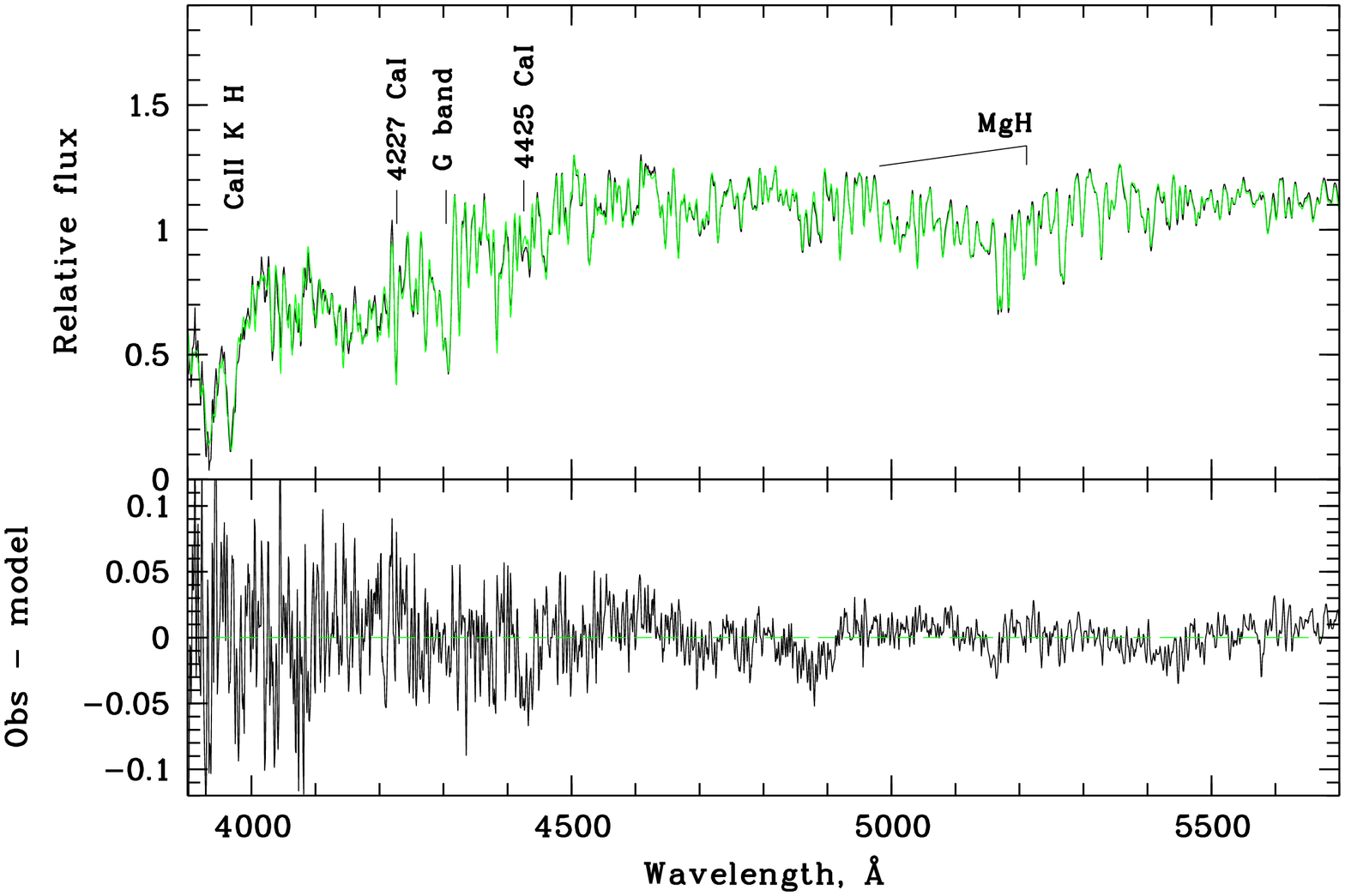}}
\resizebox{\hsize}{!}{\includegraphics[angle=-90]{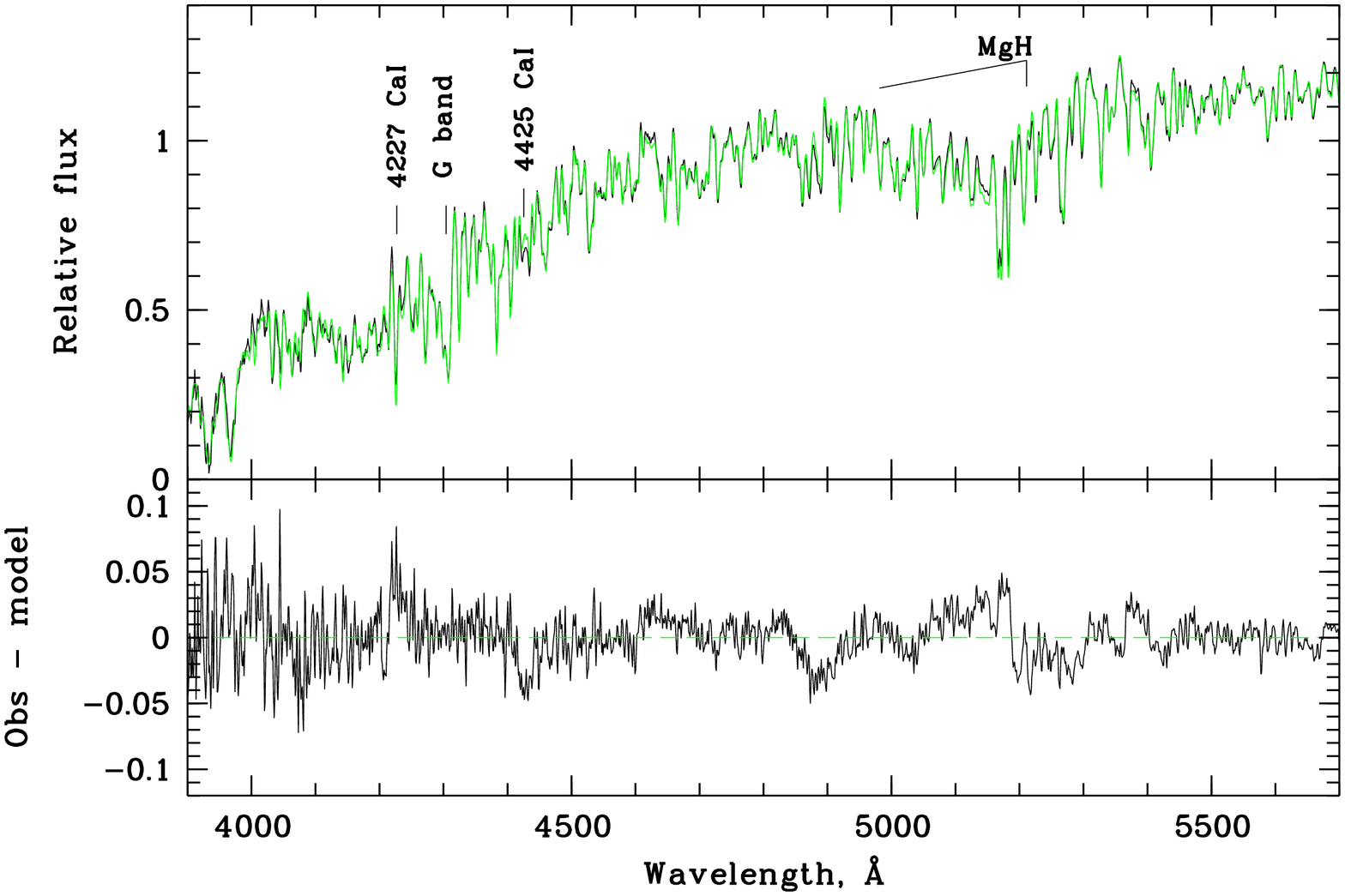}}
 \caption{Top: Fitting a representative summed spectrum of RC stars (black) using the 
{\sc ULySS} program and ELODIE 3.2 spectral library. The model spectrum is shown in grey 
(green in the electronic version).
Bottom: Comparison of the spectrum of RC stars (black) with the one of HD184406 from the MILES library 
(green in the electronic version).
The differences between the object and model spectra are shown in the lower panels, where
the dotted lines indicate zero differences.}
 \label{fig:sum}
\end{figure}

The range of parameters for the RC stars is slightly narrower:
$0.8\ge (V-I)_0 \ge 1.6$, $-1.03\le M_{I}\le0.07$, $-32\ge V_h \ge 32$, 
$-0.24\ge\feh\ge0.24$, $3900\ge T_{eff}\ge4900$, and $2.2\ge$log~g$\ge 3.5$.
It was important in our analysis to select a sample of stars in BH176 with similar
physical parameters for producing a composite spectrum. 
It was stressed by D11 and in the Introduction of this paper that only RC stars represent
a statistically significant sample of bright objects with well-defined photometric
parameters. They are not only good distance benchmarks, but also indicators of the 
evolutionary status of BH176. For comparison, Paczy\'{n}ski \& Stanek (1998) 
selected RC stars from the Hipparcos catalogue (ESA 1997) according to the following
photometric parameters: $0.8\ge (V-I)_0 \ge 1.25$ and $-1.4\le M_{I}\le1.1$.
According to Girardi \& Salars (2001), 
 the luminosities of such stars in the V and I bands are sensitive to the
ages and metallicities of the stellar ensembles they belong to, 
and also to population effects, i.e. to the element abundance anomalies 
existing in these ensembles. 

A summed spectrum with a high S/N can be used to derive more accurate radial 
velocities and atmospheric parameters for the SC. In order to estimate the 
stability of the determined characteristics we varied the number of co-added objects.
We obtain the following typical accuracies for the atmospheric parameters by comparing
the data for stars observed twice (slit 28a = slit 32b), or objects common with
the list of Frinchaboy et al. (2006) (slit 37a = F4190, slit 18a = F9312, slit 21a = F825): 
 $ \Delta \feh = 0.1 \div 0.2$ dex, $ \Delta T_{eff} = 30 \div 50$ K, 
$\Delta log(g) = 0.1 \div 0.2$ dex. 
 
The mean values for the 31 RC member stars within $\sim 3\arcmin$ form the centre of BH176  
(Table~\ref{tab:summary}) are  the following: 
$V_h= 0 \pm 15$ km/s, $ [Fe/H] = 0.1 \pm 0.1$ dex, $T_{eff}=4500 \pm 150$ K,
$log~g = 2.5 \pm 0.25$  cm/s$^2$.
The following parameters were obtained by fitting the co-added RC spectrum 
(Fig.~\ref{fig:sum}, upper panel) with the {\sc ULySS} program and the 
ELODIE 3.2 stellar library: $V_h= 4.4 \pm0.5$ km/s, $ [Fe/H] = 0.1 \pm 0.01$ dex, 
$T_{eff}=4494 \pm6$ K, and $log~g=2.56 \pm0.02$ cm/s$^2$.
The location of all the observed stars and the member ones on the CMD 
is given in Sec.~5. 

\subsection{Abundance analysis using stellar atmosphere models}
\label{sec_modatm}

 We used T$_{eff}$, $log~g$, and $\feh$ derived in the previous section for a representative
spectrum of RC stars to generate synthetic spectra using the Castelli \& Kurucz (2003) 
stellar atmosphere models and a software package originally published by Shimansky et al. (2003).
 Elemental abundance determination at medium resolution is not a new technique 
(e.g. Kirby et al. 2008 and references therein). The atmospheric parameters and abundances
can be adjusted in pixel-to-pixel spectral fits of many metal absorption lines over 
a wide spectral range. Because our spectrum is of medium resolution, 
we derived approximate abundances only for the chemical elements influencing the whole spectrum (Fe),
a large part of it (Mg, Ca, C) or having several strong dominant lines (see Table~\ref{tab:abund}).
For example, the blend of Mn ($\lambda \sim 4033 \AA$) (Fig.~\ref{fig_sam}) includes about 
twenty strong MnI lines.
The method was tested by comparing a high-resolution spectrum of Arcturus with
 the computed synthetic spectrum (Fig.~\ref{fig:Arctur}). 
The elemental abundances and atmospheric parameters for Arcturus were taken from Ram\'irez \& Prieto (2011).
We extracted the high S/N spectrum
of Arcturus ($S/N>500$) from the ELODIE library\footnote{http://atlas.obs-hp.fr/elodie/}.
The description of the method and the list of used lines and atomic constants 
(oscillator strengths, damping of wings) are given in Appendix~\ref{sec_synthetic}.

We fitted the pseudo-continuum in {\sc ULySS}. We thus transformed the shape and mean intensity 
in the observed spectrum to fit the theoretical one. When using this procedure, it is important
that the calculation of the global spectral curvature be performed exactly
 in the same way for the observed and synthetic spectra, smoothed to the observational resolution.
The observed data should have high S/N,  and
they should be cleaned and reduced properly prior to determining the pseudo-continuum.
 
The comparison between the model and the normalised observed spectrum
is shown in Fig.~\ref{fig_sam} for the most prominent spectroscopic features 
and in Fig.~\ref{fig_samall} for the whole spectrum.
One notices a strict repeatability of strong absorption line 
features consisting of individual lines and line blends in the model 
and observed spectra. The theoretical spectrum generated using 
$\feh=-0.15$ dex, $\xi_{turb}=1.6$ km$s^{-1}$ and the atmospheric parameters 
$T_{eff}=4494$~K and $log~g=2.56$ matches the  RC summed spectrum well with 
an accuracy better than 2\%\footnote{This means that the difference
between the normalised spectra is $\le0.02$}. 
The chemical compositions and their uncertainties
are summarised in Table~\ref{tab:abund}. According to these data
the $\alpha$-element abundance (mean of [Mg/Fe] and [Ca/Fe]) is 
\afe$= 0.25\pm 0.07$ dex.

\subsection{Comparison with objects in stellar libraries}

 We used the derived atmospheric parameters and $\afe$ to find analogues 
of our composite spectrum of RC stars in BH176 in the MILES 
(S\'anchez-Bl\'azquez et al. 2006, Cenarro et al. 2007) and 
ELODIE (Prugniel \& Soubiran 2001, Prugniel et al. 2007) libraries.
They both have an optimal coverage of the HR diagram. 
Nevertheless, the ELODIE coverage is not optimal for RGB stars with
metallicities $ -0.4 > \feh > 0.2$~dex (Fig. 1 in Katz et al. 2011; 
Coelho, 2009). In comparison to ELODIE, MILES represents a richer 
collection of giant stars with known atmospheric parameters and $\mgfe$ ratios.
 Milone (2011) determined Mg abundances for a large number of its objects. 
The coverage in atmospheric parameters and chemical 
patterns of MILES is large. The library serves as a starting point
for stellar population models (Vazdekis et al., 2010) and for the development 
of theoretical stellar libraries with $\alpha$-element enhanced mixtures 
(see Coelho 2014 and references therein). 

We found only one star that is similar to our object in all parameters.
This is the variable star HD~184406 (19:34:05.354 +07:22:44.18 (J2000.0)) with
$T_{eff}=4450$~K, $log~g=2.47$, $\feh=-0.13$~dex, and $\xi_{turb}=1.8$ (McWilliam, 1990).
The comparison of its spectrum with the composite spectrum of RC stars is shown
in the lower panel of Fig.~\ref{fig:sum}. 
We used a high-resolution high $S/N\sim 90$ spectrum of HD~184406 from the MILES
database\footnote{http://miles.iac.es/pages/stellar-libraries/miles-library.php}.
The comparison was done in {\sc ULySS}.
Figure~\ref{fig:sum} shows that the difference between our spectrum 
and the reference spectrum does not exceed 5\%. The abundances of chemical elements
for HD~184406 determined using the high-resolution spectra by 
Luck \& Challener (1995) and McWilliam (1990) are listed in Table~\ref{tab:abund}
for comparison with the approximate data for our representative medium-resolution spectrum.
McWilliam (1990) calibrated T$_{eff}$ using the Johnson filters and estimated 
the surface gravities from the stellar luminosities and temperatures.
We used $[C/Fe]$ and $[N/Fe]$ estimated for physical gravities. We calculated [El/Fe] from 
[M/H] values in Luck \& Challener (1995) (from Tables~11 and 6 in that paper).
If there were abundances determined for the same element from different lines,
we averaged these estimates. 
These authors derived $\feh = -0.05$, $T_{eff}=4375$~K and $log~g=2.65$ for HD~184406,
where T$_{eff}$ was determined using DDO, Geneva, and Johnson filters.
McWilliam (1990) provided $\feh$ relative to the Sun and logarithmic numerical densities of atoms.
We calculated $[El/Fe]$ abundances from these data  taking the atomic data for the Sun
from Grevesse (1984) into account as done by that author.

Table~\ref{tab:abund} shows that the $\feh$ values 
and the abundances of C, N, Mg, Ca, and Co of the two objects agree well. 
There are large differences in the literature estimates for Ca, Ti, and Ni. 
The Lick indices (Worthey 1994, Worthey \& Ottaviani 1997) measured 
in the spectra of our object and of HD~184406, smoothed to 
the resolution of $FWHM=5 \AA$, are in excellent agreement (Table~\ref{lickind2}).

\subsection{Three M-type giants}

\begin{figure}[h!]
\resizebox{\hsize}{!}{\includegraphics[width=4.5cm,angle=-90,bb=65 55 520 850]{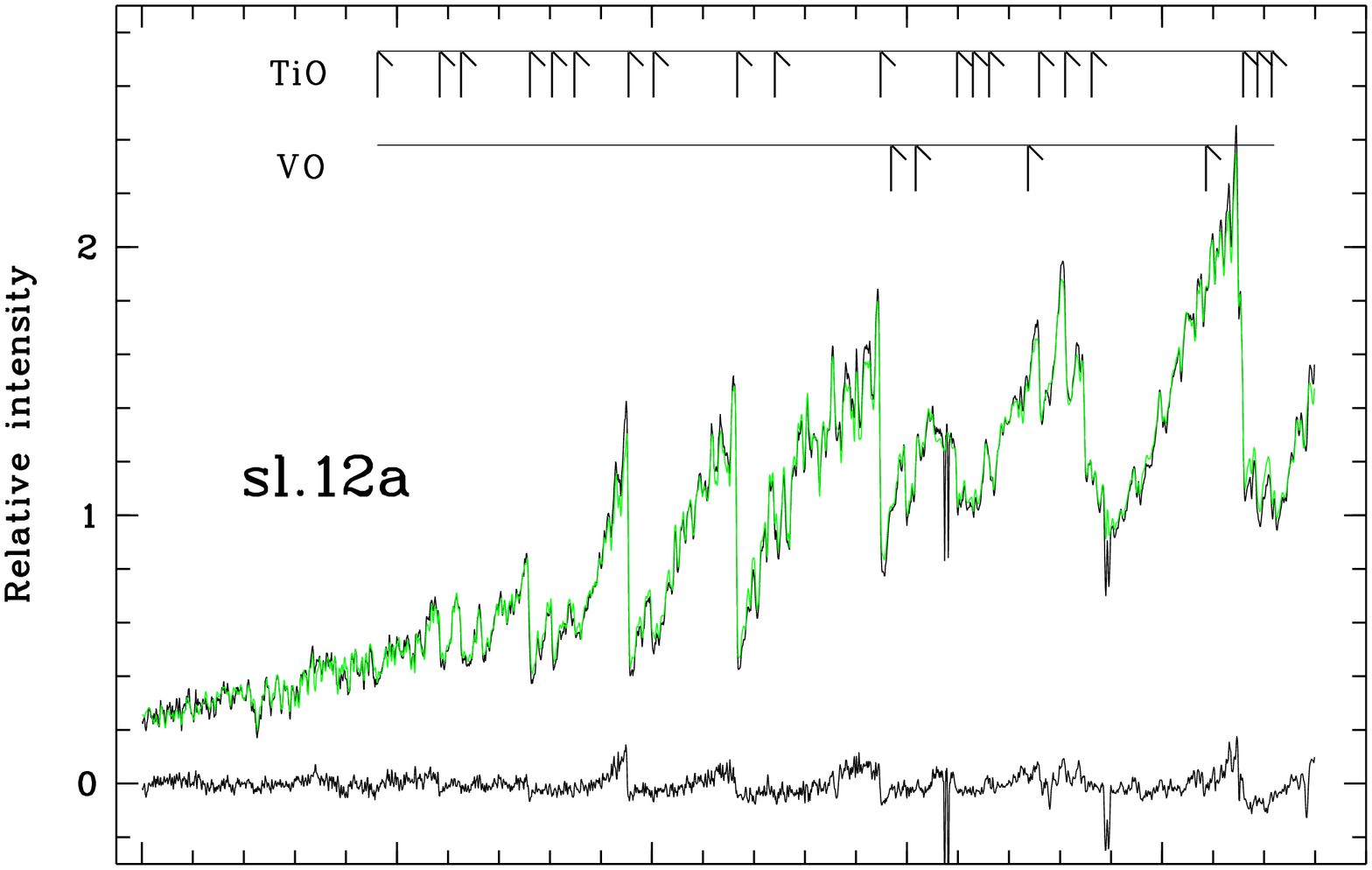}} 
\resizebox{\hsize}{!}{\includegraphics[width=4.5cm,angle=-90,bb=65 55 520 850]{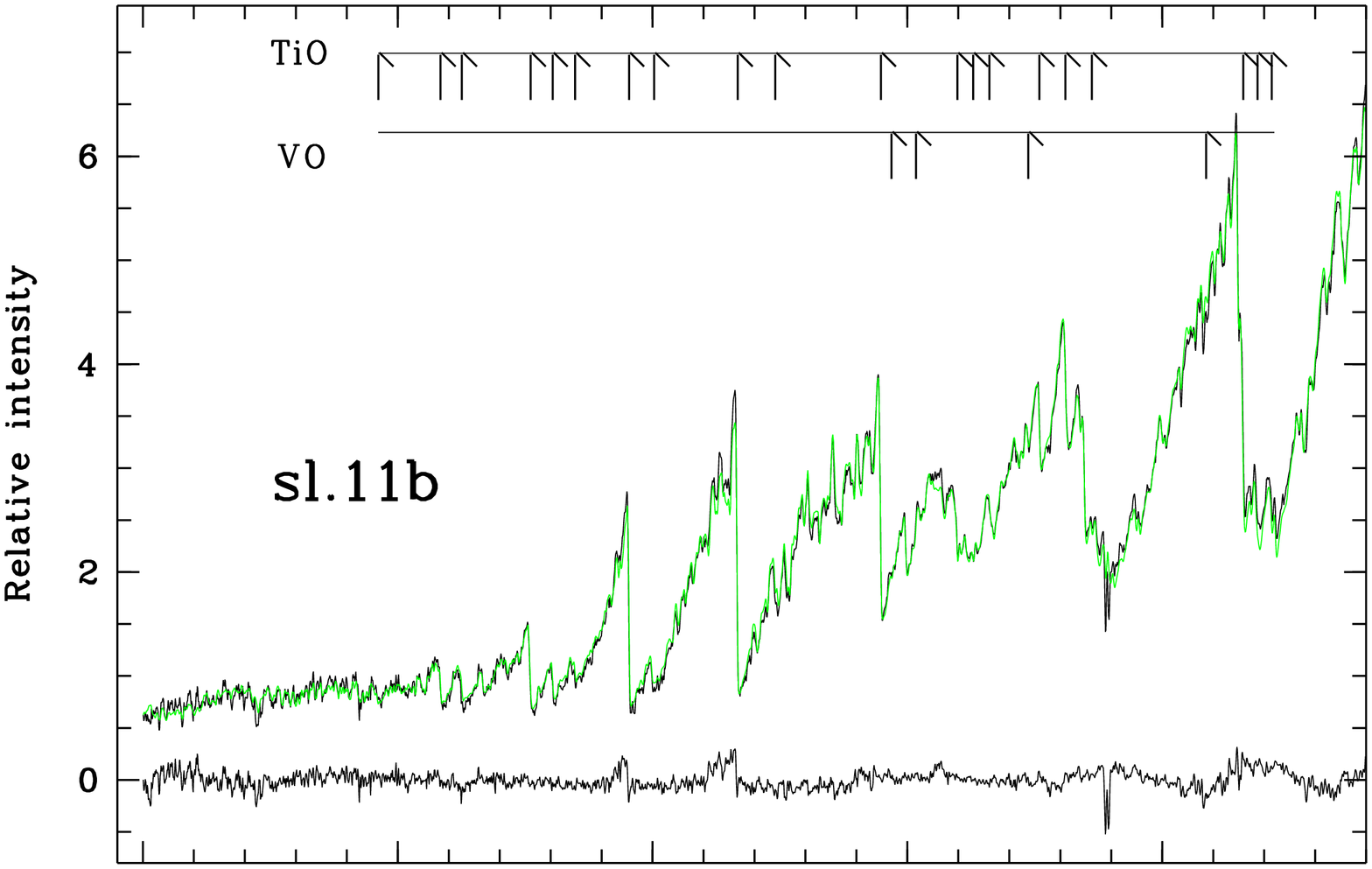}}
\resizebox{\hsize}{!}{\includegraphics[width=4.5cm,angle=-90,bb=65 55 520 850]{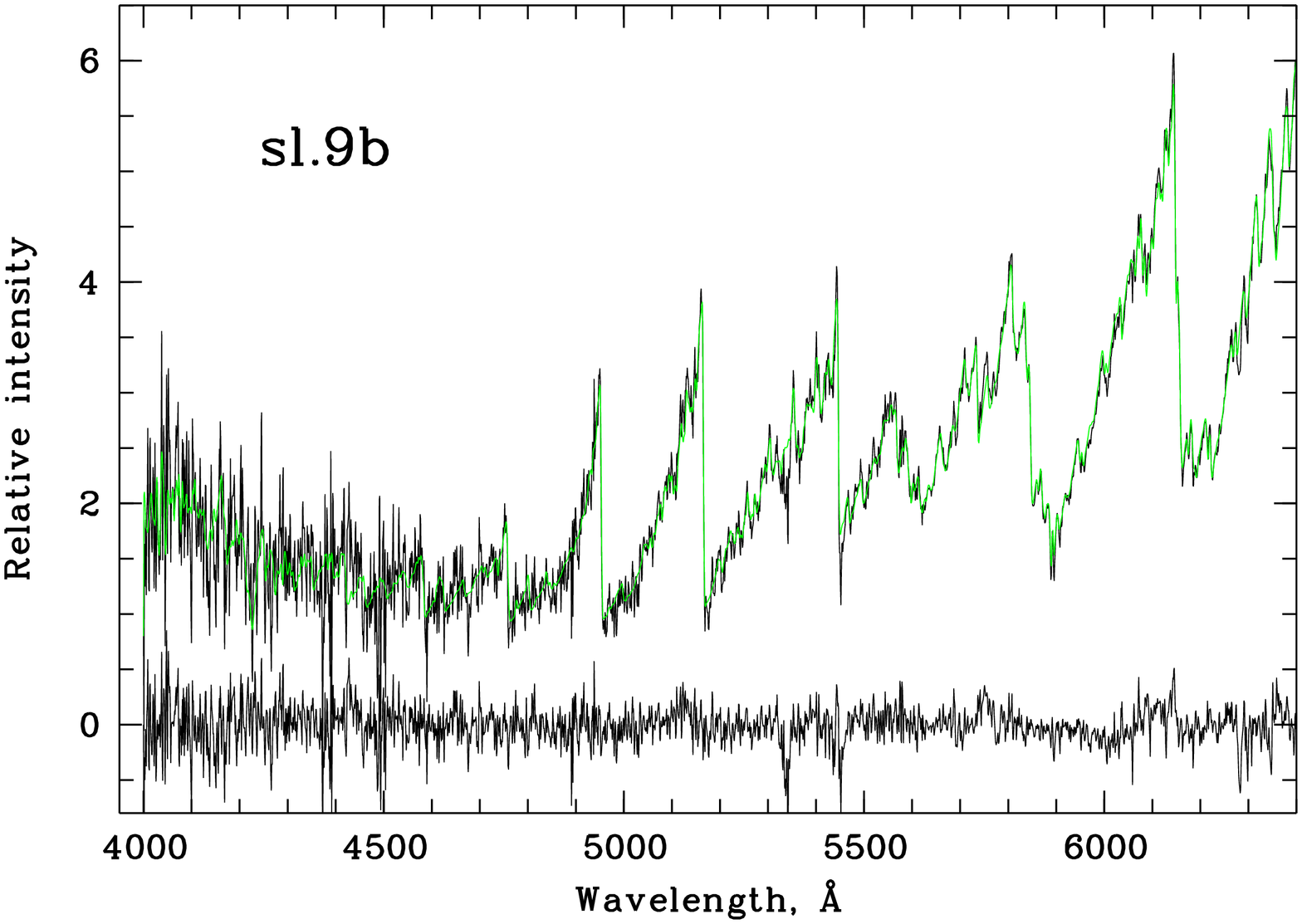}}
 \caption{Spectra of the three M-giant stars (black), slit 12a , slit 11b, and slit 9b, 
fitted with spectra of HD101153, HD169931, and HD126327, respectively (green in the electronic version).  
The differences between the object and model spectra are shown at the bottom.
The molecular bands of titanium oxide and vanadium oxide are indicated.
The derived parameters are listed in Table~\ref{tab:m_type}.}
 \label{fig:Mstars}
\end{figure} 
\begin{table*}[!hbt]
\caption{Colours and heliocentric radial velocities for the three M-type stars
and literature data for their analogues.}
\label{tab:m_type}
\begin{center}
\begin{tabular}{|lccc|ccccrccrc|}
\hline\hline 
Slit     & $V-K$ & $J-K$  & $V_h$  & Lit. analogue & $V-K$ & $J-K$  & T$_{eff}$ & $log(g)$ & $[Fe/H]$ & V ampl.& Period & Source \\
         & (mag)  & (mag)& (km $s^{-1}$)&               & (mag)  & (mag) &   (K)    & ($cm s^{-2}$) & (dex)  & (mag)  & (d)  &  \\
\hline
   12a   & 5.4   & 1.07  & 30      & HD101153      & 5.450 & 1.16   &   3452     & 0.80      & -0.08  & 0.28 & 30  &   1 \\
  11b    & 6.9   & 1.29  & 18      & HD169931      & 5.935 & 1.08   &   3106     & -0.47    &  -0.47  & 0.30 & 129 &  2 \\
 9b      & 8.6   & 1.32  & 8       & HD126327      & 9.849 & 1.26   &   3100     & 1.98     &  -0.45  & 0.20 & 157 &  2 \\
\hline
\end{tabular}
\tablebib{
(1) Percy et al. (2001); (2) Lebzelter \& Hinkle (2002).}
\end{center}
\end{table*} 

The knowledge of the radial velocities and physical properties for the three brightest
and reddest stars located within $1.5\arcmin$ from the centre of the SC 
(slits 12a, 11b, and 9b in Table~\ref{tab:summary}) is
critical for understanding the physical parameters and the nature of BH176. 
Unfortunately, we did not manage to estimate these data exhaustively using 
 the ELODIE3.2 interpolating 
arrow\footnote{ http://www.obs.u-bordeaux1.fr/m2a/soubiran/elodie$_{-}$library.html } 
because the spectral library contains very few such objects.
The fitting results are listed in Table~\ref{tab:summary}. 
The quality of the fits was poor, especially in the case of the reddest star 9b.
To better constrain the properties of the three stars, we inspected the existing spectral libraries 
for the presence of giant cool stars with similar colours and luminosities 
(Table~\ref{tab:m_type}). We also used the “Atlas of digital spectra 
of cool stars“ (Turnshek et al., 1985) to constrain the spectral class of each star 
by visual comparison.

All the selected candidate analogues were M-type giants with luminosity class III.
Such cool giant stars are fairly rare in our Galaxy and, in particular, 
in the region of the SC, judging from
the Swope spectroscopy of several hundred M giant candidates selected from the 
2MASS photometry by Majewski et al. (2004). 
The Besan\c{c}on model (see Sec~3.1) demonstrates that only $\sim0.1$\%  
stars with the same colours in this area may be giants at the same distance as BH176.

It appears that there are indeed only a few dozen M5III -- M8III stars
in the SIMBAD database\footnote{http://simbad.u-strasbg.fr/simbad/}
for which all the characteristics (colours, effective temperatures, 
surface gravities, and metallicities) are known.
Spectra of these stars are available in the ELODIE  and MILES   
stellar libraries.

We found only one analogue for each of the stars 11b and 9b (Table~\ref{tab:m_type}).
For slit 12a we found five stars with similar colours and a
full set of data from the literature.
The parameters for these five objects are in the range  $ \feh \sim -0.02 \div -0.34$,
$T_{eff} \sim 3181 \div 3452$ K and $ log(g) \sim 0.47 \div 0.8$. 
The data for three  Galactic field M giants are given in Table~\ref{tab:m_type}, 
together with the velocities and colours of our three M stars.
The columns contain the following information:
(1) slit number;  (2,3) $V-K$, $J-K$ colour; (4) heliocentric radial velocity, measured with 
respect to the reference stars;
(5) reference M-giant star; (6,7) its $V-K$ and $J-K$ colours; (8) effective temperature;  
(9) surface gravity; (10) $ [Fe/H]$; (11) amplitude of the visual variability; 
(12) period of variability; (13) literature source for the last two columns.
The  comparison between the observed and literature spectra are 
shown in Fig.~\ref{fig:Mstars}. All three  Galactic field M-stars appear to be 
giant semi-regular variables (SRb) with an amplitude of variability 
$\sim 0.3$~mag and variability periods of 30 to 160 days. 
The calculated heliocentric radial velocities allow us to conclude that our three stars belong to BH176. 
The velocities are higher than zero unlike the majority of surrounding 
Galactic stars (Fig.~\ref{fig:Vh_Fe_H}). The  T$_{eff}$ and $ log(g)$ of the three stars allow 
us to classify them as semi-regular variable giants M5III -- M8III close to the tip of the RGB.
\begin{figure*}[!hbt]
\begin{center}
\includegraphics[width=9cm, angle=-90]{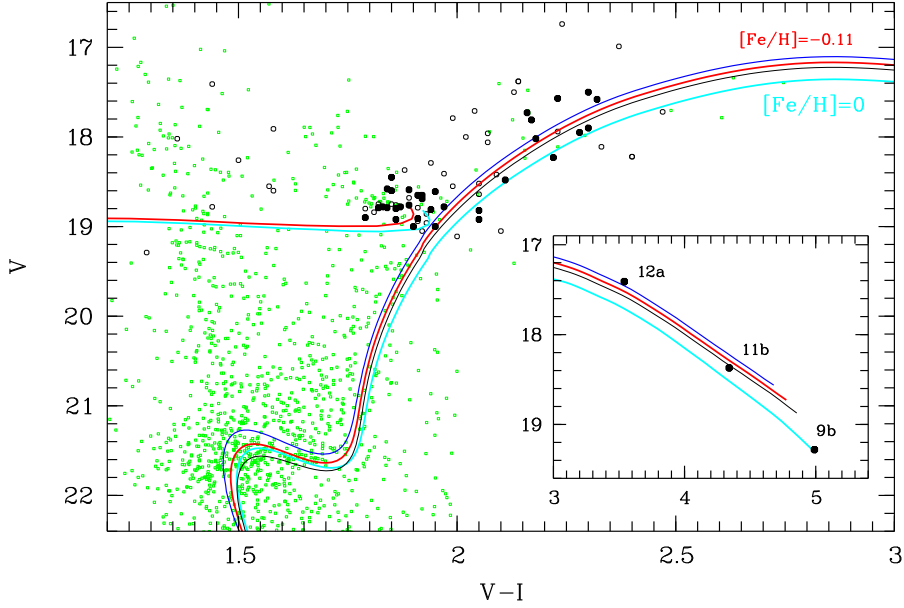}
 \caption{CMD for stars within $1.5 \arcmin$ from the centre of BH176 
(D11) (small grey dots). 
Black open squares indicate the objects observed spectroscopically.
Black filled squares show the RC and RGB stars selected according to our criteria 
to produce the co-added spectra.
Alpha-element-enhanced ($\afe=0.3$) isochrones computed from the Victoria-Regina 
stellar models of VandenBerg et al. (2006) are over-plotted: (from top to bottom)
$\feh =-0.1$ dex and Age=6 Gyr, 7 Gyr (red in the electronic version), 8 Gyr; $\feh =0.0$ dex,
Age=7 Gyr (cyan in the electronic version).}
\end{center}
 \label{fig:cmd}
\end{figure*}

\section{Proper motions}

The proper motions (PMs) in right ascension $\mu_{\alpha} cos(\delta)$ and declination $ \mu_{\delta}$ 
extracted for the program stars from the PPMXL catalogue (Roeser et al. 2010) 
are listed in Table~\ref{tab:summary} in mas yr$^{-1}$.
To identify the stars we used the on-line Aladin sky atlas\footnote{ http://aladin.u-strasbg.fr}
and coordinate transformations from the positions of the slits in the Gemini image (Sec. 2.1) 
in equatorial coordinates for epoch 2011.1 to the coordinate system of the PPMXL catalogue.
The accuracy of the latter is $\sim$80 mas relative to the $Hipparcos$ reference frame 
for stars with $K < 14$ (Roeser et al. 2010). To derive the transformation equations, we applied
the {\sc ccmap} and {\sc cctran} routines in IRAF to five bright relatively isolated stars with $K < 12$ mag.
The solution was obtained with accuracies of 0.002 and 0.037 \arcsec\ for the right ascension and 
declination, respectively. The accuracy of the solution allowed us  to calculate
PMs for some of the detected stars by comparison of the coordinates
of stars in the the PPMXL catalogue, in the VLT, and in the Gemini images. According to the mean  
$\mu_{\alpha} cos(\delta)=7 \pm 10$ mas/yr and $ \mu_{\delta}=-0.4 \pm 9$ mas/yr, BH176 
is a system unlikely to belong to the Sagittarius stream (Law \& Majewski 2010).

\section{CMD analysis}

Using the metallicity and $\alpha$-element enhancement
 for the RC stars,  $\feh=-0.15 \pm 0.1$ dex, and $\afe=0.28$ dex, obtained in Section~3,  
and the photometric results of D11,
we have re-estimated the age, distance from the Sun, and Galactic extinction for BH176
by fitting the CMD with isochrones computed from the Victoria-Regina stellar 
models of VandenBerg et al. (2006). We used the isochrone corresponding to $\feh=-0.1$ dex and $\afe=0.3$.
The photometric data and the models are shown in Fig.~\ref{fig:cmd}.
The estimated parameters for BH176 are the following: age $7\pm0.5$ Gyr, distance modulus
$(m-M)_0=15.91\pm 0.03$~mag. (distance from the Sun $15.2 \pm 0.2$~kpc), 
colour excess $E(V-I)=0.79 \pm 0.03$. The agreement with the results of D11 is very good everywhere.
Figure~\ref{fig:cmd} shows that the isochrone corresponding to the derived metallicity,
age, and \afe\ fits the observed CMD particularly well. The inset shows the red part of the CMD with 
the reddest RGB stars. The scatter of the RGB and RC stars agrees with the artificial star 
photometric results in D11. 
The isochrone $\afe=0.3$ dex, $\feh=0.0$ dex
is slightly off the mean locus of the RGB. Only the TRGB star 9b corresponds to it ``ideally''.
What might be the reason? First, this star is fainter than the two neighbours. So, its
photometric uncertainty the V-band is larger. Second, it is probable that this star is variable.
Variability is a common property of SRb stars (see Table~\ref{tab:m_type}). 
Finally, it may have chemical anomalies.

The error budget mostly includes a random component consisting of the uncertainties due to 
a variable extinction at low Galactic latitudes and of photometric errors. 
The main source of photometric errors is crowding in dense stellar fields, followed
by uncertainties in determining the background level.
The artificial star experiments (D11) have reduced the influence not only of the random 
errors on the results, but also of the systematic ones.  
The systematic difference between the intrinsic and observed colours and magnitudes includes 
unknown aperture corrections in dense stellar fields and
possible systematic shifts in the photometric zeropoints owing 
to unexpected changes in the observing conditions.
The last component may only be identified by comparing different observations of the same object.
We compared the V magnitudes for the three RGB stars (12a, 11b and 9b) using the Gemini and the VLT images
and found that 9b is brighter by $\sim$0.1 mag. in the Gemini frame. Transformations to the Johnson-Cousins
system were made using the equation of Fukugita et al. (1996).
Including the spectroscopic methods in the analysis allowed us 
to avoid the systematic differences between the observed and model colours and magnitudes 
due to unknown helium and light-element contents (see e.g. Salaris 2012 for a
detailed discussion of these issues).
 
\section{Discussion of the nature of BH176}
\label{sec_discussion}

BH176 is situated at a large Galactocentric radius: $R_{gc}=9.4$~kpc.
Its distance components in kpc relative to the Galactic centre are
$X_{gc}=-4.9$, $ Y_{gc}=-7.9$, and $ Z_{gc}=1.15$.
It is located in the outer disk according to our present-day knowledge
about the structure of the Galaxy (Gaensler et al. 2008, Moni Bidin et al. 2012, 
Adibekyan et al. 2012, Haywood et al. 2013 and references therein).
The properties of stellar populations in this area are not well known
because of its remoteness and the enormous interstellar extinction near the Galactic plane.

It cannot be a member of the Galactic anticentre stellar structure
or of the Sagittarius dwarf spheroidal galaxy, because of its too high metallicity
(Frinchaboy et al. 2006, Layden \& Sarajedini, 2000). 
BH176 is not an ordinary GC, even though it is
away from the Galactic plane and massive ($M_V = -4.2$~mag., $Radius=13.2$ pc).
Galactic GCs with metallicities $\feh\sim0$~dex are very rare (Harris 1996).
Most of them are in the bulge.
The closest GC analogue of BH176 is probably Palomar~10 with 
$R_{gc}=6.4$ kpc and distance components in kpc: $X_{gc}=4.4$, 
$ Y_{gc}=4.7$, $ Z_{gc}= 0.3$ (Harris 1996). 
Its metallicity is $\feh \sim -0.1$ dex on the Zinn-West metallicity scale 
(Kaisler et al. 1997) and $ M_V=-5.8$ mag. The origin of Palomar~10 has not
been firmly established.
 
BH176 is too metal-rich to be one of the OCs associated with the 
outer Galactic disk (Frinchaboy et al. 2006, Frinchaboy \& Majewski 2008,
 Villanova et al. 2010).
The mean metallicity of the outer-disk OCs and field stars 
is $\feh \sim -0.48 \pm 0.12$~dex (Bensby et al. 2011, Jacobson et al. 2011),
and  $\afe \sim -0.2$ dex (L\'epine et al., 2014).
The metallicity and age of BH176 is similar to that of the 
thin-disk field stars, but the $\alpha$-element content is higher 
than for the thick disk component with $\feh \sim 0$
(see e.g. Figs.~7 -- 9 in Haywood et al., 2013).  
Only a significant contribution of very massive stars 
that end their lives as SNII can give rise to such an enrichment.
However, if the thick disk finished forming stars $\sim8$ Gyr ago
(Haywood et al., 2013),
the gas density in this area $\sim7$ Gyr ago was not as high as  
in the thin disk, where OCs were effectively destroyed.

We selected OCs having ages greater than 4 Gyr from the catalogue of
Gozha, Borkova, and Marsakov (2012) (hereafter: GBM12).
Table~\ref{tab_oc} in the Appendix summarises the  
data for the resulting sixteen objects.
 The successive columns list the following information: (1) OC name;
(2) absolute visual magnitude, corrected for Galactic extinction; (3) colour excess;
(4) age; (5) \feh; (6) \mgfe;  (7) Galactocentric radius;  (8) heliocentric distance towards the north
 Galactic pole; (9) distance from the Sun. The data in columns 1 and 2 were calculated by us 
using the SIMBAD and NASA/IPAC Extragalactic Database (NED). All other data are from Gozha et al. (2012).

The mean characteristics are the following:  
$\feh=-0.14 \pm 0.28$~dex, $\mgfe=0.16 \pm 0.1$~dex, $Age = 5.8 \pm 1.8$~Gyr.
These 16 OCs reside at Galactocentric distances in the range 6.6 -- 20.3 kpc,
and heliocentric distances toward the north Galactic pole from
$Z=-2.51$~kpc (Berkeley 20) and -1.77~kpc (Berkeley 75) to $Z=1.1$~kpc (NGC6791) 
and 1.7~kpc (Saurer~1) (Table~\ref{tab_oc}). Eleven OCs of the sixteen 
are located above $\vert Z\vert \sim 0.5$ kpc. Their absolute 
visual magnitudes are in the range $M_{V_0} \sim -4.9 \div -1.1$~mag.
Nine of the selected clusters were classified by GBM12 as belonging to the 
thick disk. Some of them may belong to the outer disk as well. 
For example, Carraro et al. (2007) considered Berkeley 25 and 75 to be members of the outer disk. 
There may be other, probably even younger, analogues of such objects that have $\feh \sim0$ and high $\afe$ and
are inclined with respect to the Galactic surface orbits (see e.g. Friel et al. 2010, Glushkova et al. 2009).
 BH176 is evidently a member of this old OC sample according to its properties.

Another question is what the nature of these presumably thick disk OCs is.
That the $\alpha$-element ratios are high, the ages are in a very
wide range, and the locations under the Galactic plane are very different may mean that such 
objects may have originated as embedded massive SCs from the material of the thin disk,
perhaps as the result of its dense gas layers interaction
with high-velocity clouds or dark mini-haloes. A
similar explanation was suggested by GBM12. These objects
were not destroyed by the dissipative processes in the Galactic disk, 
because their orbits were strongly inclined with respect to its surface.

Part of these clusters might have appeared because of the interaction of the thin disk
with gas-rich dwarf satellites moving through it.
The recent discovery of vertical waves in the Galactic disk is evidence fof the passage of a small galaxy 
or dark matter sub-halo  (Widrow et al. 2012). 
There is an extragalactic example of such an interaction. 
The bright spiral galaxy NGC~6946 located at a distance of $\sim 5.9$~Mpc (Karachentsev et al. 2000) 
has experienced recent powerful star formation bursts revealed by a number of supernovae 
and star-forming regions.

One of them contains a super star cluster that is as bright as a dwarf galaxy
(see Efremov et al. 2007 and references therein).
The origin of this object was interpreted by the impact of a high-velocity gas super cloud.
Post-impact shock wave collisions are usually associated with high gas pressure, which
favours the formation of massive SCs (Elmegreen \& Efremov, 1997).

\section{Summary}

Medium-resolution observations of stars in BH176 using the Gemini South Multi-Object Spectrograph, 
combined with deep VLT photometry, allowed us to determine the distance,
chemical composition, and evolutionary status of this outer-disk SC more accurately. 
The candidate stars for spectroscopic observation were selected using 
the photometric results of D11.
Thirty-one RC stars were identified in BH176 with radial velocities, photometric,
and atmospheric parameters in the range
$1.8\ge V-I \ge 2.4$, $16.\ge I\ge17.1$, $-32\ge V_h \ge 32$, 
$-0.24\ge\feh\ge0.24$, $3900\ge T_{eff}\ge4900$, $2.2\ge log~g\ge 3.5$.
The identification of RC stars was indispensable for deriving the distance and extinction
for the SC correctly, because only these objects are numerous
and bright enough in BH176 to provide a reliable benchmark. Identifying the RC stars 
with atmospheric parameters in a narrow range
was also important for computing a co-added spectrum that was bright and accurate enough
to be approximated by the model of one star.
Full-spectrum fitting methods were used to estimate radial velocities, metallicities,
effective temperatures, and surface gravities for the individual stars.
Stellar atmosphere models and comparison with individual stars in 
the MILES and ELODIE stellar libraries 
were used to derive the same parameters plus abundances 
of different elements for a co-added high S/N spectrum of RC stars.
Finally, the age, distance from the Sun, and Galactic extinction were specified
by comparison of the photometric results in the V and I bands with theoretical
isochrones.
The fundamental parameters of three bright red M-type giant near the TGRB were
estimated using literature analogues with similar spectra and 
broad-band optical and infrared colours.

 The properties of BH176 are fully consistent with the ones
of Galactic OCs older than 4 Gyr presumably belonging to the thick disk (GBM12).
We speculate that BH176 may have originated as a super-star cluster 
as a consequence of encounter of the forming thin disk with a high-velocity cloud
or with a gas-rich dwarf galaxy. BH176 is an interesting target for future
deep photometric and high-resolution spectroscopic studies with large telescopes.

\begin{acknowledgements}
The Gemini data for this paper were obtained under programme GS-2011A-Q-22
of the Argentinian time share. This research made 
use of the NASA Astrophysics Data System Bibliographic services (ADS), the 
NASA/IPAC Extragalactic Database (NED)\footnote{The NASA/IPAC Extragalactic Database (NED) is operated 
by the Jet Propulsion Laboratory, California Institute of Technology, under contract 
with the National Aeronautics and Space Administration.}, the SIMBAD database operated at the CDS 
(Strasbourg, France), and Google. SME acknowledges a grant RFBR14-02-96501-r-ug-a.
SVV acknowledges a grant RFBR13-02-00351. We thank Vladimir Marsakov for useful discussions, 
which helped us to improve the paper, and an anonymous referee for comments.
\end{acknowledgements}



\clearpage
\appendix
\section{Computation of synthetic spectra using stellar atmosphere models}
 \label{sec_synthetic}

To derive the abundances of different chemical elements, we computed synthetic
 spectra with and without atomic and molecular lines.
We used the software package {\sc SPECTR} (Shimansky et al. 2003, 
see also Menzhevitski et al. 2014 for its latest version). 
Shimanskaya et al. (2011) used this program for deriving chemical abundances for 
 the secondary in the close binary system FF Aqr. They show that 
lines of chemical elements in high-resolution spectra strongly broadened by rotation
are reproduced well by the theoretical spectra. 

Our synthetic spectra are based on plane-parallel, hydrostatic stellar atmosphere models
for a given  set of parameters (T$_{eff}$, $log~g$, $[M/H]$) 
computed by interpolating a model grid of Castelli \& Kurucz (2003)
using the technique described by Suleimanov (1996). The solar chemical abundances 
and isotope compositions were specified using the data of 
Asplund et al. (2006) for Fe, C, N, and O and
of Anders and Grevesse (1989) for all the other elements.

The radiative transfer equation at each frequency (wavelength)
was solved using the Hermite method with the determination
of specific radiation intensities for fixed angles. 
For a model atmosphere, we calculated the
fluxes emerging in three basic
directions, with inclinations to the surface of $62\degr$,
$30\degr$, $8\degr$. The stellar surface was then subdivided into
sectors. The fluxes emerging from sectors towards the
observer were derived by interpolating the radiation intensities 
at the three basic angles to the actual visibility angle. 

The computation of synthetic spectra took 
about 600 000 atomic and more than 1 800 000 $ ^{12}CH$,
$ ^{13}CH$, and $SiH$ molecular lines into account  from the lists of
Kurucz (1994) and Castelli \& Kurucz (2003), and 28
bands of 10 molecules (VO, TiO, SO, SiO, NO, MgO, MgH, CO, CN, AlO) 
computed in terms of the theory of Nersisyan et al. (1989) 
and kindly provided by Ya.V. Pavlenko. 
We used the empirical oscillator strengths 
from Shimanskaya et al. (2011) for 1350 strong optical lines in the 
$ \lambda 3900–7000 \AA$ wavelength range.
   
We computed the profiles of the HI lines according to the broadening 
theories of Vidal, Cooper \& Smith (1973) and Griem (1960). 
The standard Voigt profiles for the remaining lines were calculated with broadening
due to thermal motion and microturbulence, natural damping, 
Stark broadening in the approximation of Kurucz and Furenlid (1979), 
and van der Waals broadening with constants determined
from the classical Gray formula (Uns\"{o}ld, 1955) with scaling
factors $\Delta log C_{6} = 0.7\div1.2$ (Shimanskaya et al. 2011).
The synthetic spectra were binned to the resolution of 0.05\AA.
We estimated the intensity uncertainties of the line profiles provided by this value 
to be within 0.005\% of the continuum flux.

We modelled the line profiles  HI, MgI, MgII, AlI, and CaII taking
deviations from local thermodynamic equilibrium (LTE) into account. For each model
atmosphere, the non-LTE populations were computed
using the complete linearisation method of Auer \& Heasley (1976) 
with the package NONLTE3 of Sakhibullin (1983).

Summing the radiation from all the aforementioned sectors of the stellar 
atmosphere considering their areas and local radial 
velocities due to stellar rotation and radial and tangential 
microturbulence yielded the integrated radiation from the atmosphere. 
The resulting stellar spectra were broadened in accordance with the instrumental
function of the spectrograph.

\subsection{Determining the chemical abundances}
\label{sec_sam}

Since the full width at half maximum in our observed spectra is $FWHM\sim 5\AA$,
almost all spectral lines are weak and blended. The most intense of them are dominant.
At this resolution, it is possible to derive the abundance of a chemical element accurately
($\sigma \le0.15$ dex) in spectra with high S/N per resolution bin ($S/N \ge 100$),
if this element 1) influences the whole spectrum 
or a large part of it ($\Delta \lambda \sim 100 \AA$), and 2) has several strong dominant lines.
If an element has just one dominant line and contributes $ \ge 50\%$ of the line intensity,
the accuracy of the derived abundance is $\sim 0.2$ dex.
The iron abundance and the microturbulent velocity ($\xi_{turb}$) influence the whole spectrum.
If $\xi_{turb}$ is incorrect, it is not possible to achieve an optimal agreement
between strong and weak iron lines in the theoretical and observed spectra.
The macroturbulent and stellar rotation velocities were not taken into account,
because these effects are too weak to be detectable at our observational resolution
(Smith \& Dominy 1979, Gray \& Toner 1986).
The following elements have prominent atomic absorption lines that are dominant at our resolution:
MgI 5167\AA, 5172\AA, 5183\AA, CaI 4226\AA, CaII 3933\AA,\ and CaII 3968\AA.
The shapes of strong molecular bands, such as CH, MgH including hundreds of lines. 
are easily recognised and reproduced well by the synthetic spectrum. 

Figure~\ref{fig_sam} shows theoretical spectra calculated with the stellar atmosphere model
corresponding to the following parameters:
 $[M/H]=0.0$ dex, $T_{eff}=4494$~K, and log$g=2.56$.
In each of the panels, abundances for one of the elements are varied, while the abundances of other
elements have the fixed values listed in Table~\ref{tab:abund}.
The derivation of the iron abundance is illustrated in the top right-hand panel.
In the course of fitting \feh, the micro-turbulent velocity $\xi_{turb}$ was estimated
(see top left panel of Fig.~\ref{fig_sam}). These two values (\feh\ and $\xi_{turb}$) are closely 
related. 
A normalised summed spectrum of 31 RC stars (Table~\ref{tab:summary}) is shown by a solid line.
The fitting of the whole spectrum is illustrated in Fig.~\ref{fig_samall}.
\begin{figure*}
\begin{tabular}{p{0.5\textwidth}p{0.5\textwidth}}
\hspace{-1.5cm}
\includegraphics[height=8.3cm,width=11.5cm]{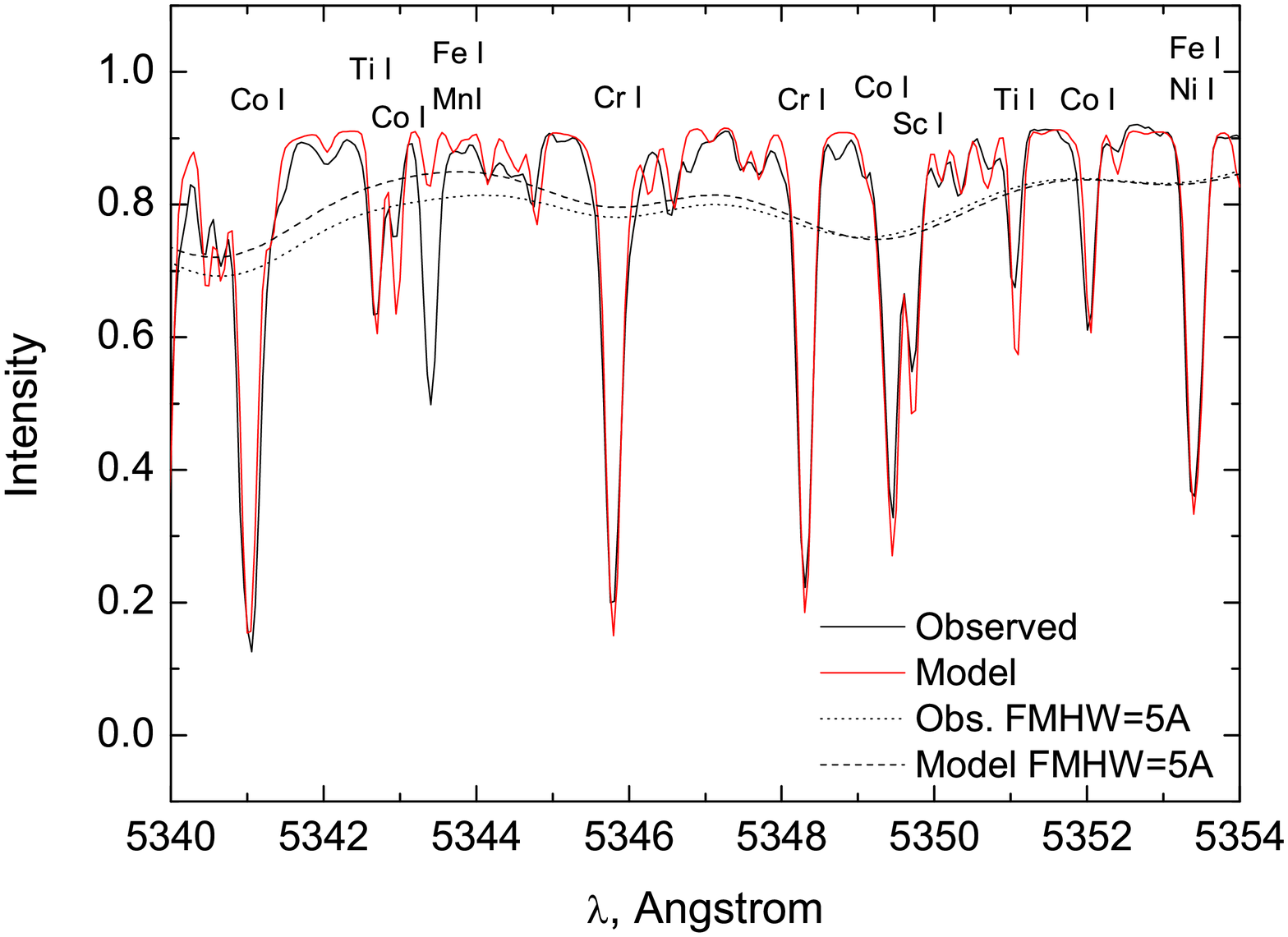} &
\hspace{-1.0cm}
\includegraphics[height=7.8cm,width=10.cm]{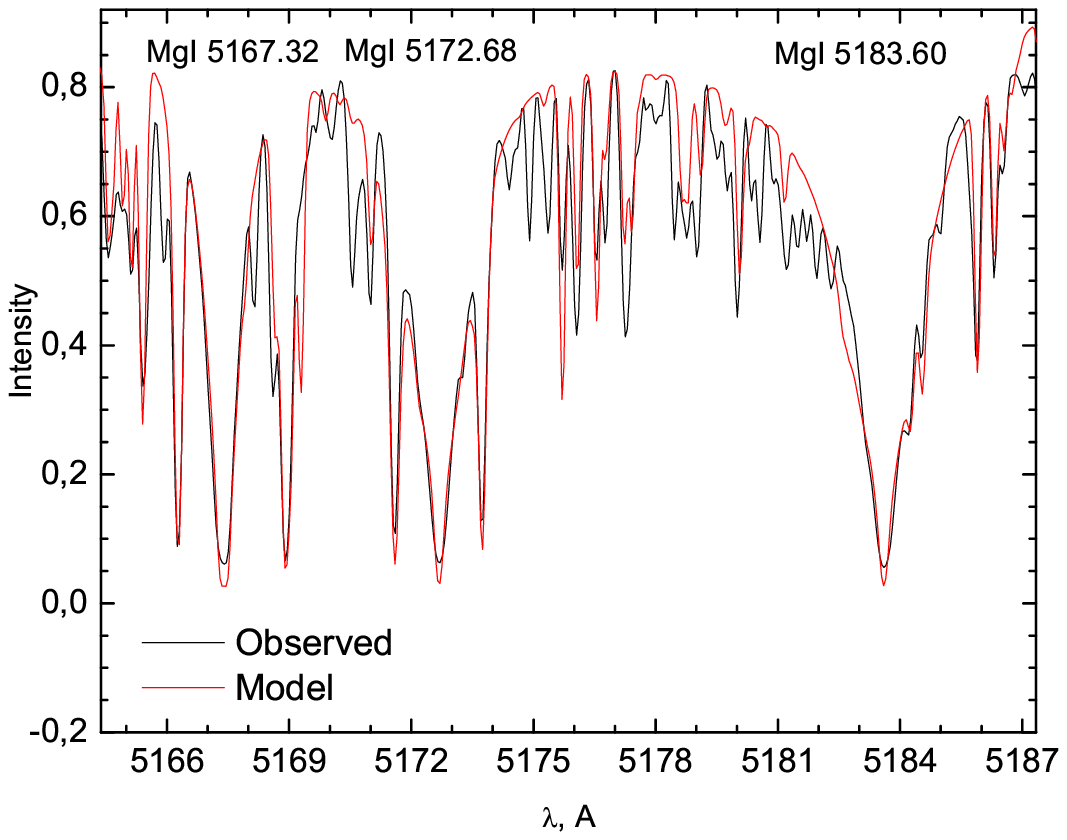} \\
\hspace{-1.3cm}
\includegraphics[height=8.2cm,width=10.5cm]{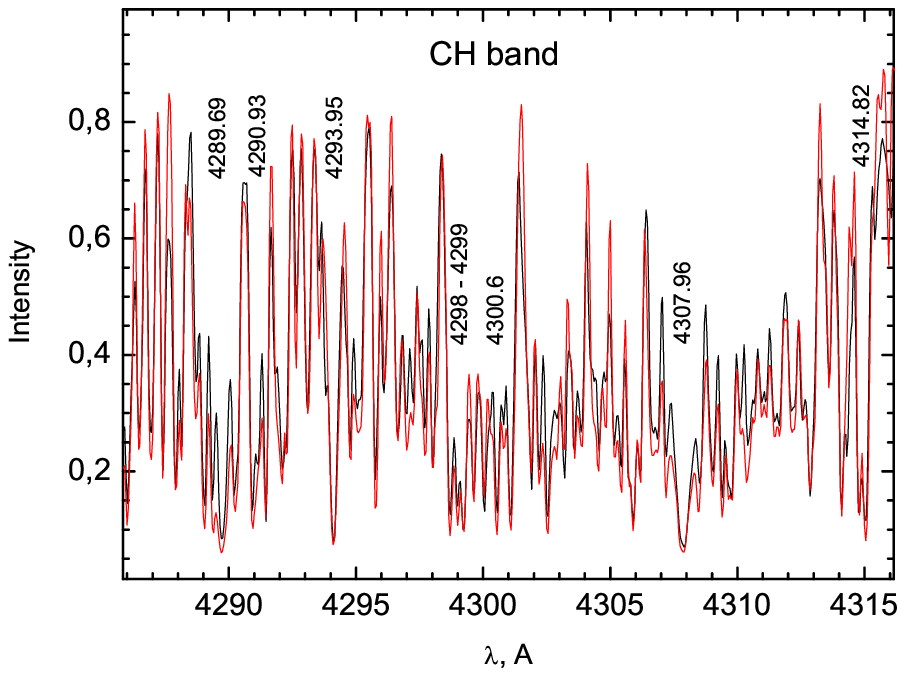} &
\hspace{-1.5cm}
\includegraphics[height=8.2cm,width=10.4cm]{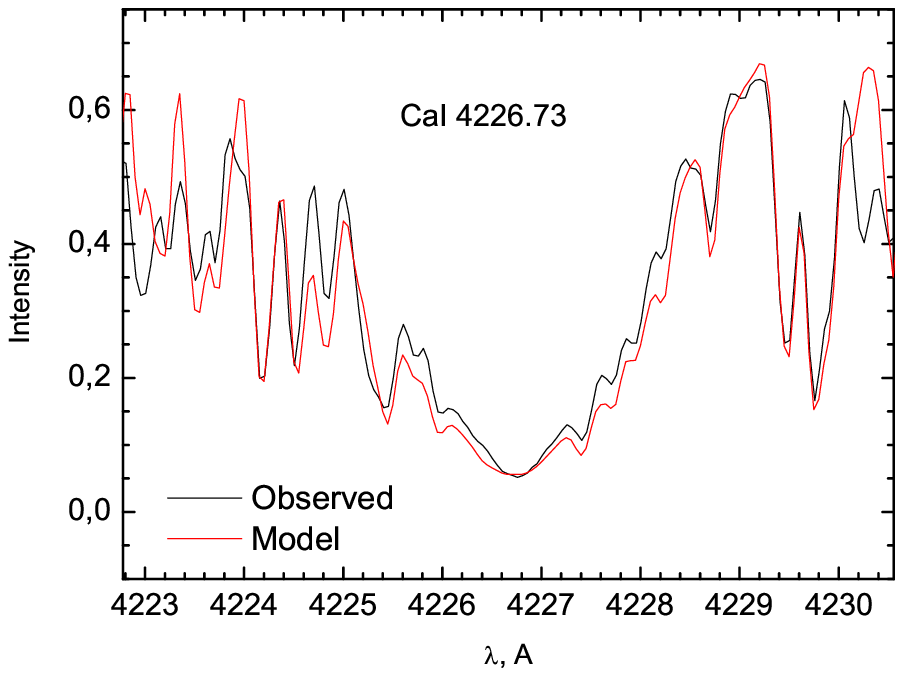} \\
\end{tabular}
 \caption{Comparison of the observed Arcturus spectrum from the ELODIE database and our calculated synthetic spectrum 
for a wavelength region including iron peak elements and for three
wavelength regions corresponding to several prominent spectroscopic features (CH and MgH molecules, CaII H, K lines, CaI 4227\AA). 
The adopted parameters are from Ram\'irez \& Prieto (2011).}
 \label{fig:Arctur}
\end{figure*} 

\onecolumn
\begin{figure*}[!h]
\includegraphics[width=19cm, height=23cm]{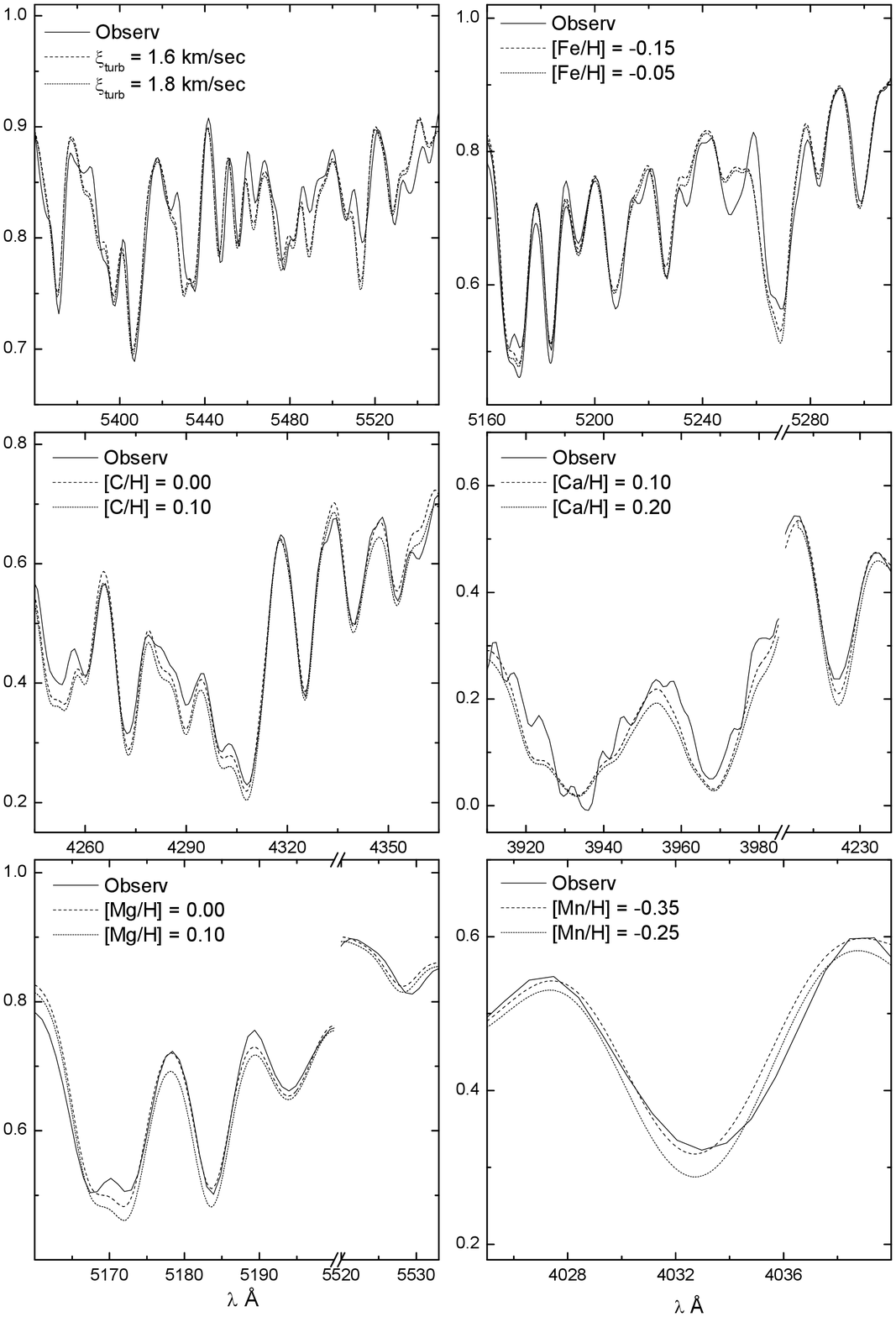}
 \caption{Illustration of fitting the abundances of different chemical elements
(Section~3.2). The derived abundances are listed in Table~\ref{tab:abund}.}
 \label{fig_sam}
\end{figure*}
\begin{figure*}[!h]
\includegraphics[width=19cm, height=23cm]{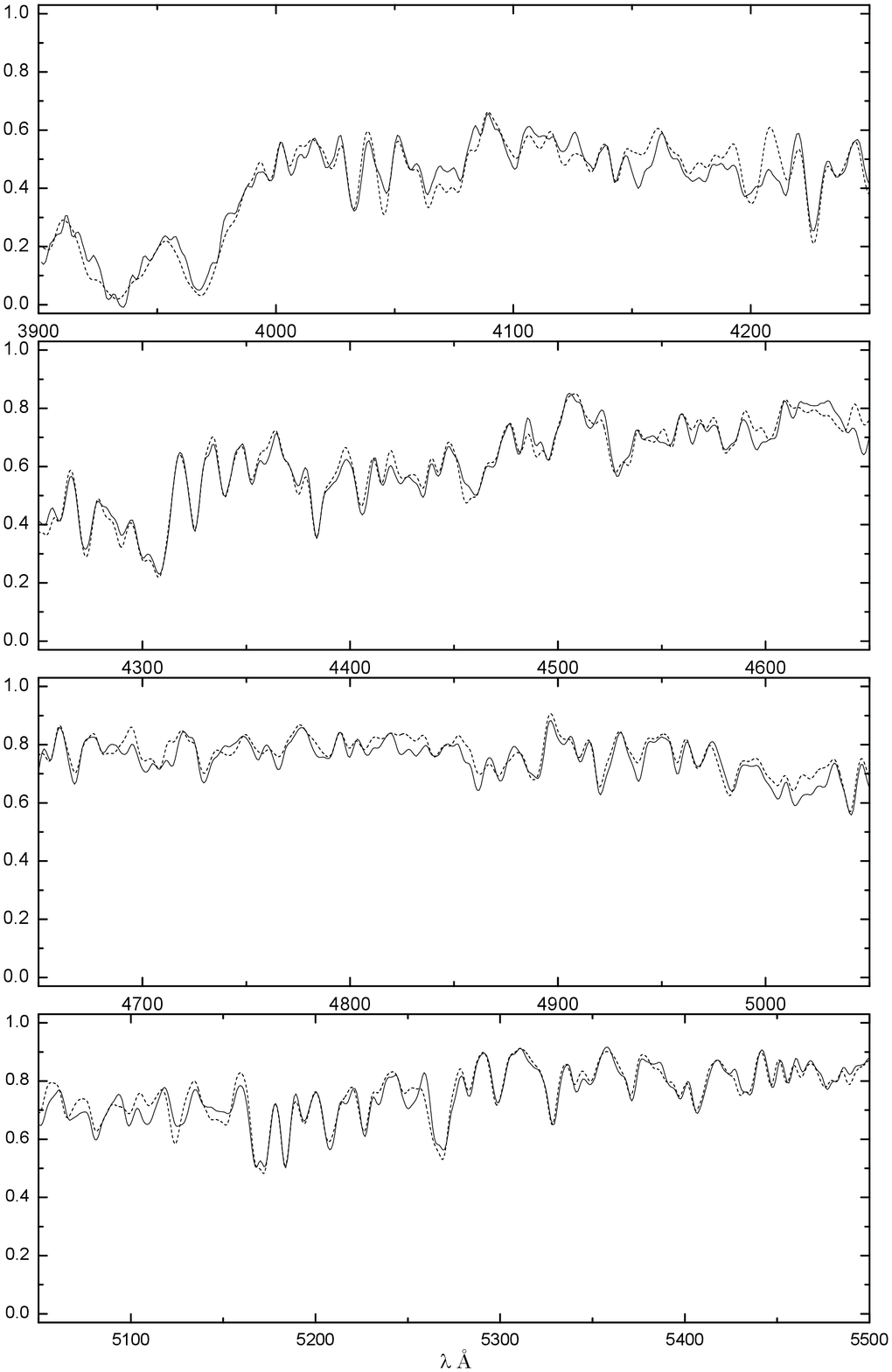}
\caption{Fitting the whole spectrum of the RC stars in BH176 by the synthetic one (see Appendix~\ref{sec_synthetic}) 
The derived abundances are listed in Table~\ref{tab:abund}.}
\label{fig_samall}
\end{figure*}
\newpage
\begin{table*}[!hbt]
\caption{Parameters for all stars observed spectroscopically (see Sec.~\ref{sec_ulyss} for details).}
\label{tab:summary}
\vspace{-0.7cm}
\scriptsize
\begin{center}
\begin{tabular}{rcccccccrcccccrl}
\\ \hline\hline\noalign{\smallskip}
Slit& RA (J2000.0) DEC        & $I$ &  $V-I$ & $K_s$ & $J-K_s$ & $T_{eff}^{V-K}$ &  Dist. & $V_h$       &  $ \feh$ & T$_{eff}$ & $log~g$  &  $\mu_{\alpha}cos(\delta)$ & $ \mu_{\delta}$  & $ S/N$ & Note \\     
    & ($ h$ \hspace{0.1cm}$ m$ \hspace{0.1cm} $ s$ \hspace{0.4cm} $ ^{o}$ \hspace{0.1cm} $ \arcmin$ \hspace{0.1cm} $ \arcsec$) & {\it(mag)} & {\it(mag)}  & {\it(mag)} & {\it(mag)} &  {\it(K)}& ($ \arcsec$)& {\it(km/s)}& {\it(dex)} & {\it(K)}& ($cm/s^2$)&   {\it(mas/yr)}& {\it(mas/yr)}&   &  \\    \hline \noalign{\smallskip}
01a & 15:39:23.835-50:04:18.53& 15.99 &  2.07& 13.00 & 0.85    &  4051        &  166.0 & 32$\pm$10  &   0.40  &   4170  & 2.6  &   40           &    29     & 39 & {\it RGB?}   \\
02a & 15:39:24.756-50:03:31.22& 17.02 &  1.58&       &         &              &  153.1 & -63$\pm$5 &   0.22  &   5870  & 4.2  &                &           & 26 &        \\
03a & 15:39:22.423-50:01:14.75& 16.33 &  1.58&       &         &              &  155.3 & -74$\pm$11 &  -0.27  &   4854  & 2.8  &                &           & 32 &        \\
04a & 15:39:22.960-50:00:37.45& 16.76 &  1.50& 14.88 & 0.60    &   5583       &  183.9 & -62$\pm$5 &   0.18  &   5639  & 4.5  &                &           & 30 &        \\
05a & 15:39:20.099-50:02:37.23& 14.62 &  2.37& 11.65 & 0.93    &   3900       &  103.3 & -33$\pm$5 &   0.02  &   3893  & 1.4  &      -7        &    -1     & 56 & $^{**}${\it RGB,TiO}    \\
06a & 15:39:20.685-50:00:08.75& 16.49 &  1.88& 14.16 & 0.65    &   4568       &  192.7 & -161$\pm$5&  -0.68  &   4467  & 2.0  &       4        &    5      & 24 &        \\
07a & 15:39:18.376-50:04:47.50& 16.98 &  1.57& 14.73 & 0.72    &    4977      &  147.5 & -9$\pm$15  &  -0.06  &   5100  & 3.5  &                &           & 28 & {\it RC?} \\
08a & 15:39:12.019-50:02:47.41& 16.60 &  1.85&       &         &              &   24.9 & 32$\pm$7  &   0.08  &   4635  & 2.3  &      19        &   11      & 32 & $^*${\it RC}     \\
09a & 15:39:15.627-50:02:36.85& 15.97 &  1.44&       &         &              &   59.8 & -5$\pm$10  &   0.03  &   6000  & 4.1  &                &           & 56 &        \\
10a & 15:39:11.242-50:03:14.81& 16.74 &  1.91& 14.68 & 0.72    &   4808       &   32.0 & -32$\pm$13 &  -0.24  &   4852  & 3.5  &                &           & 34 & $^*${\it RC?}     \\
11a & 15:39:08.698-50:03:04.50& 16.74 &  1.84&       &         &              &   18.5 & -2$\pm$14  &   0.15  &   4491  & 2.7  &                &           & 28 & $^*${\it RC}     \\
12a & 15:39:07.829-50:03:10.43& 13.87 &  3.54& 10.24 & 1.05    &     3519     &   27.3 &  30:       &   -0.02 &   3261  & 0.6  &      12        &     8     & 20 & {\it SRb}    \\
13a & 15:39:06.999-50:03:16.63& 16.80 &  1.97&       &         &              &   37.0 & 18$\pm$15  &   0.22  &   4596  & 2.6  &      -5        &   -4      & 23 & $^*${\it RC}     \\
14a & 15:39:09.249-50:02:50.23& 16.73 &  1.92&       &         &              &    2.9 & -11$\pm$18 &   0.07  &   4297  & 2.3  &      14        &    1      & 24 & $^*${\it RC}     \\
15a & 15:39:13.199-50:03:35.23& 15.34 &  2.23& 12.50 & 0.84    &   4035       &   59.6 & 23$\pm$12  &   0.18  &   4019  & 2.1  &      9         &    5      & 44 & $^{**}${\it RGB}   \\
16a & 15:39:17.168-50:04:18.90& 17.13 &  1.92& 14.53 & 0.75    &    4348      &  118.2 & 24$\pm$18  &   0.34  &   4428  & 2.6  &                &           & 20 & {\it RC?}       \\
17a & 15:39:14.725-50:00:41.88& 17.11 &  2.00& 14.65 & 0.61    &   4403       &  136.1 & 5$\pm$12   &   0.20  &   4409  & 2.7  &                &           & 17 & $^*${\it RC}     \\
18a & 15:39:12.595-50:01:59.00& 15.67 &  2.28& 12.86 & 0.83    &   4020       &   57.6 & 0$\pm$15   &   0.12  &   3973  & 1.9  &      5         &    -6     & 37 & $^{**}${\it RGB,TiO}   \\
19a & 15:39:13.732-50:03:34.03& 16.47 &  2.05& 13.69 & 0.73    &              &   61.9 & -64$\pm$5 &  -0.43  &   4931  & 3.4  &                &           & 51 &        \\
20a & 15:39:16.379-50:03:09.73& 16.87 &  2.05& 14.10 & 0.90    &   4166       &   70.3 &  6$\pm$17  &   0.18  &   4286  & 2.5  &      -3        &     3     & 21 & $^*${\it RC}     \\
21a & 15:39:10.609-50:02:27.41& 16.91 &  1.87&       &         &              &   23.4 & 8$\pm$18   &   0.06  &   4508  & 2.2  &      7         &    -14    & 19 & $^*${\it RC}     \\
22a & 15:39:10.060-50:02:22.50& 16.91 &  1.86&       &         &              &   25.5 & -59$\pm$5 &  -0.28  &   5512  & 4.2  &                &           & 45 &        \\
23a & 15:39:17.685-50:01:08.08& 15.80 &  1.99&       &         &              &  127.6 & 13$\pm$18  &  -0.83  &   3878  & 4.6  &                &           & 35 & {\it Mg-rich?}      \\
24a & 15:38:56.585-50:05:06.38& 15.82 &  2.40& 12.93 & 0.90    &   3923       &  186.1 & 20$\pm$12  &  0.12   &   3966  & 2.2  &    5           &    -11    & 34 & {\it RGB?}       \\
25a & 15:39:05.899-50:03:26.43& 16.01 &  2.22& 13.22 & 0.74    &   4063       &   51.4 & 20$\pm$11  &   0.21  &   4031  & 2.3  &                &           & 31 & $^{**}${\it RGB}   \\
26a & 15:39:03.110-50:03:03.94& 18.00 &  1.29&       &         &              &   64.3 & -96$\pm$9 &  -0.59  &   6006  & 4.4  &                &           & 24 &        \\
27a & 15:39:00.618-50:03:30.10& 17.02 &  1.79&       &         &              &   94.9 & 1$\pm$10   &   0.28  &   4554  & 2.7  &                &           & 39 & $^*${\it RC}     \\
28a & 15:38:58.105-50:01:33.20& 16.67 &  1.95&       &         &              &  132.5 & 16$\pm$15  &   0.02  &   4283  & 2.4  &                &           & 25 & $^*${\it RC}     \\
29a & 15:38:57.099-50:02:48.90& 15.89 &  2.07& 13.29 & 0.97    &              &  119.0 & 15$\pm$11  &   0.08  &   3955  & 2.2  &     19         &    11     & 34 & {\it RC?}    \\
30a & 15:38:55.819-50:01:09.80& 15.25 &  2.47& 12.28 & 0.94    &   3871       &  164.2 & -10$\pm$7 &   0.19  &   3822  & 1.4  &    6           &    -1     & 39 & $^{**}${\it RGB}    \\
31a & 15:39:01.959-50:03:43.43& 15.37 &  2.13& 12.49 & 0.89    &    4057      &   91.0 & 21$\pm$5  &   0.00  &   3950  & 2.2  &                &           & 51 & {\it RC?}    \\
32a & 15:39:03.929-50:02:51.70& 15.84 &  2.18& 13.11 & 0.87    &   4113       &   53.2 & 7$\pm$19   &   0.07  &   4086  & 2.0  &     15         &    14     & 35 & $^{**}${\it RGB,TiO}    \\
33a & 15:38:59.419-50:01:52.40& 16.88 &  1.91&       &         &              &  111.5 & 1$\pm$18   &   0.15  &   4524  & 2.5  &                &           & 27 & $^*${\it RC}     \\
34a & 15:39:04.908-50:01:46.80& 16.87 &  1.89& 14.10 & 1.09    &   4258       &   75.5 & -4$\pm$20  &   0.02  &   4442  & 2.5  &      -4        &    -13    & 25 & $^*${\it RC}     \\
35a & 15:39:06.535-50:01:38.41& 16.60 &  2.05& 14.35 & 0.79    &    4497      &   75.1 & -7$\pm$20  &  -0.86  &   3807  & 4.6  &                &           & 21 & {\it Mg-rich?}       \\
36a & 15:39:02.478-50:04:59.10& 15.71 &  2.23& 12.88 & 0.77    &   4037       &  147.7 & 14$\pm$15  &   0.12  &   4067  & 2.3  &                &           & 41 & {\it RGB}     \\
37a & 15:38:59.908-50:04:13.70& 15.24 &  2.14&       &         &              &  125.8 & 15$\pm$5  &   0.17  &   4194  & 2.5  &                &           & 59 & {\it RGB}     \\
38a & 15:39:01.375-50:01:52.31& 16.93 &  1.93&       &         &              &   95.7 & -12$\pm$20 &   0.20  &   4475  & 2.2  &                &           & 23 & $^*${\it RC}     \\
39a & 15:39:05.318-50:02:37.23& 17.09 &  1.90& 14.21 & 1.05    &   4178       &   41.1 & 14$\pm$15  &   0.01  &   4415  & 2.5  &      5         &    -8     & 25 & $^*${\it RC}     \\
40a & 15:38:54.389-50:01:36.20& 17.03 &  1.81&       &         &              &  162.0 & -28$\pm$15 &   0.08  &   4613  & 2.4  &                &           & 25 & $^*${\it RC}   \\
41a & 15:38:55.168-50:04:51.98& 15.77 &  2.33& 12.90 & 0.85    &   3968       &  185.5 & 33$\pm$18  &   0.18  &   3900  & 2.1  &      -6        &    -12    & 38 & $^*${\it RC}  \\
01b & 15:39:22.310-50:02:46.22& 15.67 &  2.04&       &         &              &  123.2 & -72$\pm$5 &   0.24  &   4328  & 2.6  &                &           & 47 &           \\
02b & 15:39:25.082-50:01:04.15& 17.03 &  1.93&       &         &              &  183.0 & -13$\pm$18 &   0.05  &   4476  & 2.1  &                &           & 20 &  $^*${\it RC}    \\
03b & 15:39:20.448-50:01:13.30& 16.36 &  1.94&       &         &              &  142.1 & 2$\pm$13   &  -1.28  &   4236  & 1.0  &                &           & 33 &        \\
04b & 15:39:19.549-50:03:16.51& 17.03 &  1.91&       &         &              &  101.6 & -4$\pm$19  &   0.14  &   4587  & 2.4  &                &           & 22 & $^*${\it RC}     \\
05b & 15:39:18.223-50:03:39.70& 16.66 &  1.36&       &         &              &   98.8 & -48$\pm$11 &   0.22  &   5802  & 4.6  &                &           & 47 &        \\
06b & 15:39:17.452-50:01:20.81& 14.50 &  2.24& 11.93 & 1.36    &  4168        &  116.5 & -20$\pm$5  &   0.05  &   4036  & 1.8  &                &           & 72 &  {\it RGB,TiO}      \\
07b & 15:39:11.878-50:04:14.78& 16.95 &  2.10&       &         &              &   90.4 & -84$\pm$18 &  -0.13  &   5819  & 4.4  &                &           & 43 &        \\
08b & 15:39:08.309-50:03:34.81& 16.76 &  1.85&  14.31& 0.68    &   4531       &   48.2 &  9$\pm$15  &   0.13  &   4603  & 2.8  &      1         &    10     & 30 & $^*${\it RC}     \\
09b & 15:39:10.599-50:03:03.75& 14.30 &  4.99&  8.89 & 1.31    &              &   19.0 & 8:        &         &         &      &       20       &     4     & 87 & {\it SRb}    \\
10b & 15:39:11.258-50:03:10.40& 16.95 &  1.84&  14.68& 0.72    &   4691       &   29.0 & -2$\pm$20  &   0.19  &   4545  & 2.4  &     -14        &    14     & 24 & $^*${\it RC}     \\
11b & 15:39:06.855-50:02:53.51& 14.03 &  4.34&  9.70 & 1.28    &              &   25.4 &  18:       &  -0.06  &   3106  & 0.0  &      15        &    4      & 30 & {\it SRb}   \\
12b & 15:39:07.409-50:03:35.83& 16.87 &  1.94&  14.77& 0.84    &   4743       &   51.8 & 8$\pm$13   &  -0.11  &   4513  & 2.5  &      22        &    7      & 27 & $^*${\it RC}     \\
13b & 15:39:05.295-50:02:43.73& 15.60 &  2.30&  12.73& 0.84    &   3982       &   40.1 & 4$\pm$11   &   0.11  &   3999  & 1.7  &      7         &   -4      & 41 & $^{**}${\it RGB,TiO}    \\
14b & 15:39:06.378-50:02:55.30& 16.94 &  1.86&       &         &              &   30.5 & -16$\pm$17 &   0.13  &   4580  & 2.4  &                &           & 21 & $^*${\it RC}     \\
15b & 15:39:09.150-50:02:20.70& 15.57 &  2.16& 12.90 & 0.77    &   4160       &   27.1 & -5$\pm$7  &   0.14  &   4093  & 2.0  &      4         &    -6     & 46 & $^{**}${\it RGB,TiO}    \\
16b & 15:39:12.478-50:04:15.20& 16.57 &  1.99& 13.99 & 0.66    &   4337       &   92.8 &  4$\pm$11  &   0.29  &   4242  & 2.7  &                &           & 29 & {\it RC?}       \\
17b & 15:39:14.059-50:03:24.86& 17.06 &  1.86& 13.69 & 0.73    &   3964       &   57.8 & -25$\pm$7  &   0.24  &   4646  & 2.8  &                &           & 23 & {\it RC?}       \\
18b & 15:39:15.089-50:03:08.03& 16.34 &  2.09& 13.72 & 0.70    &   4235       &   58.0 & -3$\pm$11  &   0.06  &   4197  & 2.3  &       -1       &     4     & 28 & $^{**}${\it RGB}    \\
19b & 15:39:14.615-50:02:20.81& 16.38 &  2.11& 13.72 & 0.91    &   4199       &   56.6 & -4$\pm$15  &   0.13  &   4143  & 2.3  &       6        &    -4     & 28 & $^{**}${\it RGB,TiO}    \\
20b & 15:39:13.592-50:02:58.73& 17.04 &  1.95&       &         &              &   41.3 & 0$\pm$17   &   0.14  &   4470  & 2.3  &      -11       &     -1    & 21 & $^*${\it RC}     \\
21b & 15:39:05.778-50:03:12.70& 15.26 &  2.32&       &         &              &   43.5 & -10$\pm$13 &   0.09  &    3864 & 1.8  &                &           & 40 & $^{**}${\it RGB}    \\
22b & 15:39:09.579-50:02:50.73& 16.77 &  1.92& 14.32 & 0.70    &   4459       &    3.0 & 6$\pm$12   &  -0.05  &   4339  & 2.4  &      6         &    0      & 26 & $^*${\it RC}     \\
23b & 15:39:16.369-50:02:49.33& 16.89 &  1.85& 14.55 & 0.71    &   4594       &   66.9 & 9$\pm$18   &   -0.17 &   4019  & 1.9  &                &           & 23 &  {\it Mg-rich?}      \\
24b & 15:39:03.749-50:03:49.83& 16.77 &  2.05& 14.43 & 0.58    &   4451       &   82.8 & -7$\pm$11  &   0.15  &   4292  & 2.6  &      25        &     2     & 25 & $^*${\it RC}     \\
25b & 15:39:03.160-50:02:43.10& 17.34 &  1.44&       &         &              &   61.6 & -3$\pm$18  &   0.18  &   5576  & 4.2  &                &           & 28 &        \\
26b & 15:39:04.718-50:04:42.71& 16.71 &  1.89& 14.44 & 0.66    &   4652       &   23.8 & -6$\pm$18  &   0.15  &   4436  & 2.9  &                &           & 32 &  $^*${\it RC}     \\
27b & 15:39:01.749-50:03:07.13& 15.64 &  2.17& 12.99 & 0.79    &   4168       &   76.6 & -2$\pm$13  &   0.19  &   4126  & 1.9  &      -1        &    -9     & 44 & $^{**}${\it RGB,TiO}   \\
28b & 15:39:00.489-50:03:34.73& 15.99 &  2.02& 13.43 & 0.73    &   4321       &   98.2 & -2$\pm$7  &   0.10  &   4178  & 2.5  &                &           & 39 & $^{**}${\it RGB}   \\
29b & 15:39:01.008-50:04:11.73& 17.11 &  1.79& 14.85 & 0.62    &   4742       &   16.8 & -10$\pm$15 &   0.08  &   4770  & 2.3  &      -2.       &     -22   & 24 & $^*${\it RC}     \\
30b & 15:39:02.498-50:02:48.63& 16.97 &  1.82& 14.19 & 1.03    &   4303       &   66.9 & -1$\pm$17  &   0.16  &   4573  & 2.5  &      26.       &    -5     & 26 & $^*${\it RC}     \\
31b & 15:38:54.032-50:00:17.88& 15.20 &  2.30& 12.38 & 0.78    &   4003       &   11.5 & -29$\pm$5 &   0.17  &   4029  & 1.7  &                &           & 45 & $^{**}${\it RGB,TiO} \\
32b & 15:38:58.105-50:01:33.20& 16.67 &  1.95& 14.53 & 0.99    &   4712       &   32.5 & -13$\pm$18 &   0.04  &   4252  & 2.4  &     6          &    -15    & 25 & $^*${\it RC}     \\
33b & 15:38:54.725-50:01:59.90& 16.93 &  1.82&       &         &              &   49.8 & -90$\pm$15 &  -0.97  &   4420  & 1.8  &                &           & 29 &        \\
34b & 15:38:55.435-50:04:19.50& 16.43 &  1.97&       &         &              &   63.1 & -49$\pm$11 &   0.53  &   6992  & 0.0  &                &           & 37 &        \\
35b & 15:38:56.078-50:02:42.50& 16.95 &  1.83& 13.94 & 0.68    &   4155       &   28.9 & -6$\pm$17  &   0.15  &   4581  & 2.4  &                &           & 24 & $^*${\it RC}     \\
36b & 15:38:57.308-50:02:59.60& 17.00 &  1.91& 14.48 & 0.57    &   4422       &   17.6 & -12$\pm$19 &   0.18  &   4574  & 2.4  &                &           & 22 & $^*${\it RC}    \\
37b & 15:38:59.348-50:03:48.40& 16.79 &  1.89&       &         &              &   14.6 & 25$\pm$18  &   0.49  &   6987  & 0.0  &      33        &    -16    & 28 &       \\
38b & 15:38:56.585-50:05:06.38& 15.82 &  2.40& 12.93 & 0.90    &   3933       &   86.1 & 27$\pm$13  &   0.13  &   3945  & 2.1  &       5        &    -11    & 41 &  {\it RC?}      \\
39b & 15:38:59.908-50:04:13.70& 15.24 &  2.14& 14.50 & 1.40    &              &   25.8 & 3$\pm$11   &   -0.03 &   4132  & 1.95 &      31        &     -1    & 63 & $^{**}${\it RGB,TiO}  \\
\noalign{\smallskip}
\hline
\end{tabular}
\end{center}
\end{table*}
\begin{table*}
\caption{Lick indices ($\lambda \le 4531$\AA\ ) (first line)
measured in the summed spectrum of RC stars in BH176 and in the MILES medium-resolution spectrum of HD184406
with uncertainties (second line indicated by the "$\pm$" sign) determined from
bootstrapping of the object spectrum.}
\label{lickind2}
\begin{tabular}{llrrrrrrrrr} \\
\hline \hline
ID            & H$\delta_{\rm A}$ & H$\gamma_{\rm A}$& H$\delta_{\rm F}$ &  H$\gamma_{\rm F}$ & CN$_1$  & CN$_2$ & Ca4227 & G4300 & Fe4383 & Ca4455    \\  
               &(\AA)             & (\AA)          &  (\AA)              &  (\AA)             & (mag)    & (mag)   & (\AA)  & (\AA) & (\AA)  & (\AA)  \\
 \hline
  RC in BH176      & -6.64  & -9.32& -1.71& -3.29& 0.242  & 0.284 & 2.05  & 6.20  & 7.96  &  2.09     \\
 \hskip 50pt $\pm$ & 0.35   & 0.32 &  0.20&  0.22& 0.008  & 0.01  & 0.22 &  0.39  & 0.30  &  0.17     \\
HD184406           & -6.26  &-9.28 & -1.65& -3.28& 0.251  & 0.293 & 2.01  &  6.24  & 7.70  &  2.15     \\
 \hskip 50pt $\pm$ & 0.37   & 0.32 &   0.2&  0.21& 0.008  & 0.010 & 0.22  &  0.41  &  0.44 &  0.17    \\
\hline  \hline
\end{tabular}
\end{table*}
\begin{table*}
\caption{Lick indices ($\lambda \ge 4531$\AA\ ) (first line)
measured in the summed spectrum of RC stars in BH176 and in the MILES medium-resolution spectrum of HD184406
with uncertainties (second line indicated by the "$\pm$" sign) determined from
bootstrapping of the object spectrum.}
\label{lickind2}
\begin{tabular}{llrrrrrrrrr} \\
\hline \hline
ID                 & Fe4531 & Fe4668 & H$\beta$ & Fe5015 & Mg$_1$   & Mg$_2$  & Mg$b$   & Fe5270 & Fe5335 & Fe5406 \\
(S/N)              & (\AA) & (\AA)  &  (\AA)   & (\AA)  & (mag)    & (mag)     & (\AA) & (\AA)  & (\AA)  & (\AA)  \\
 \hline
 RC in BH176       &  3.67  &  8.98 & 1.13& 7.06 & 0.129 & 0.270  & 3.73& 4.08& 3.75&   2.69    \\
 \hskip 50pt $\pm$ &  0.01  &  0.17 & 0.07& 0.23 &  0.002&  0.003 & 0.07& 0.08& 0.08&    0.04   \\
 HD184406          & 3.70   &  8.71 & 1.12& 7.12 & 0.146 & 0.283  & 3.57& 4.12& 3.73 & 2.69     \\
 \hskip 50pt $\pm$ & 0.11   &  0.17 & 0.03& 0.23 & 0.002 & 0.003  & 0.07& 0.08& 0.09 & 0.04     \\
\hline  \hline
\end{tabular}
\end{table*}
\begin{table*}
\caption{Open clusters older than 4 Gyr from the catalogue of Gozha et al. (2012).
See Sec.~\ref{sec_discussion} for details.}
\label{tab_oc}
\begin{tabular}{lrrrrrrrr} \\
\hline \hline
 Cluster         &  $M_{V_0}$ & $E(B-V)$ & $Age$ &  $\feh$ & $\mgfe$ & Z      & $Dist_{GC}$ & $Dist_{\sun}$     \\  
                 &  (mag)   &   (mag)    & (Gyr) & (dex) & (dex)     & (kpc)  & (kpc)       & (kpc)  \\
\hline                                                            
 NGC 6791        &  -4.95 &   0.14    & 4.4 &  0.32 & 0.13           &  1.11  &  8.17       &  5.85  \\
 NGC 188         &  -4.16 &   0.08    & 4.3 &  0.12 & 0.26           &  0.78  &  9.20       &  2.05  \\
 King 2          &        &   0.36    & 6.0 & -0.42 &                & -0.47  & 12.12       &  5.75  \\
 NGC 1193        &  -1.08 &   0.21    & 5.0 & -0.22 & 0.23           & -0.96  & 12.03       &  4.57  \\
 Berkeley 18     &        &   0.54    & 4.3 &  0.02 &                &  0.51  & 13.65       &  5.80  \\
 Berkeley 70     &        &   0.58    & 4.7 & -0.32 &                &  0.26  & 12.09       &  4.17  \\
 Berkeley 17     &        &   0.68   & 10.0 & -0.10 & 0.12           & -0.17  & 10.69       &  2.70  \\
 Trumpler 5      &  -3.83 &   0.82    & 5.0 & -0.30 &                &  0.04  & 10.25       &  2.40  \\
 Berkeley 20     &        &   0.22    & 6.0 & -0.61 & 0.22           & -2.51  & 15.88       &  8.40  \\
 Collinder 106   &        &   1.09    & 7.9 &       &                & -0.01  &  8.70       &  0.77  \\
 Saurer 1        &        &   0.15    & 5.0 & -0.38 & 0.05           &  1.70  & 20.26       & 13.20  \\
 Berkeley 39     &        &   0.12    & 7.9 & -0.21 & 0.15           &  0.84  & 11.90       &  4.78  \\
 Berkeley 25     &        &   0.40    & 5.0 & -0.20 & 0.11           & -1.92  & 17.82       & 11.40  \\
 Collinder 261   &  -2.61 &   0.40    & 8.9 & -0.03 & 0.14           & -0.21  &  7.10       &  2.19  \\
 NGC 6253        &  -1.85 &   0.32    & 5.0 &  0.43 & 0.01           & -0.17  &  6.61       &  1.58  \\
 Berkeley 75     &        &   0.10    & 4.0 & -0.22 & 0.38           & -1.77  & 15.17       &  9.10  \\
\hline \hline
\end{tabular}
\end{table*}                                              
\end{document}